%% file: main.tex
\documentclass{jpp}

\usepackage{amsmath,amssymb,amsfonts}
\usepackage{bm}
\usepackage{physics}
\usepackage{verbatim}
\usepackage{hyperref}
\usepackage{natbib}
\usepackage{placeins}
\usepackage{adjustbox}
\usepackage{tikz}
\usepackage{soul}
\usepackage{subcaption}
\usepackage{cleveref}
\usepackage{graphicx}
\usepackage{float}
\usepackage{xcolor}
\usepackage{mathtools}
\usepackage[percent]{overpic}

\crefformat{equation}{(#2#1#3)}
\Crefformat{equation}{(#2#1#3)}
\crefrangeformat{equation}{(#3#1#4)--(#5#2#6)}
\Crefrangeformat{equation}{(#3#1#4)--(#5#2#6)}
\crefmultiformat{equation}{(#2#1#3)}{ and~(#2#1#3)}{, (#2#1#3)}{ and~(#2#1#3)}
\Crefmultiformat{equation}{(#2#1#3)}{ and~(#2#1#3)}{, (#2#1#3)}{ and~(#2#1#3)}
\crefrangemultiformat{equation}{(#3#1#4)--(#5#2#6)}{ and~(#3#1#4)--(#5#2#6)}{, (#3#1#4)--(#5#2#6)}{ and~(#3#1#4)--(#5#2#6)}
\Crefrangemultiformat{equation}{(#3#1#4)--(#5#2#6)}{ and~(#3#1#4)--(#5#2#6)}{, (#3#1#4)--(#5#2#6)}{ and~(#3#1#4)--(#5#2#6)}
\crefname{figure}{figure}{figures}
\Crefname{figure}{figure}{figures}
\crefname{subfigure}{figure}{figures}
\Crefname{subfigure}{figure}{figures}
\crefname{table}{table}{tables}
\Crefname{table}{table}{tables}
\crefname{section}{$\S$}{$\S\S$}
\Crefname{section}{$\S$}{$\S\S$}
\crefname{subsection}{$\S$}{$\S\S$}
\Crefname{subsection}{$\S$}{$\S\S$}

\makeatletter
\@addtoreset{equation}{section}
\patchcmd{\@maketitle}{(Received xx; revised xx; accepted xx)\hfill}{}{}{}
\gdef\@underjournal{}
\makeatother

\definecolor{red}{rgb}{1.0,0.0,0.0}
\definecolor{gre}{rgb}{0.5,0.5,1.5}
\definecolor{blu}{rgb}{0.0,0.0,1.0}

\begin{document}

\title[Relaxed magnetohydrodynamics with cross-field flow]{Relaxed magnetohydrodynamics with cross-field flow}

\author[A. Tavassoli, S.R. Hudson, Z.S. Qu and M. Hole]{Arash Tavassoli$^1$\thanks{Email address for correspondence: arash.tavassoli@anu.edu.au},
Stuart R. Hudson$^2$, Z.S. Qu$^3$ and Matthew Hole$^1$}

\affiliation{$^1$Mathematical Sciences Institute, Australian National University, Acton, ACT 2601, Australia\\
$^2$Proxima Fusion GmbH, Bavaria 81369 Munich, Germany\\
$^3$School of Physical and Mathematical Sciences, Nanyang Technological University, Singapore 637371, Republic of Singapore}

\maketitle

\begin{abstract}
The phase-space Lagrangian model of Dewar \emph{et al.}~(Phys. Plasmas 27, 062507, 2020) provides a model for incorporating cross-field flow into relaxed equilibria while retaining ideal magnetohydrodynamics force balance. Here, we characterise the steady-state solution space and identify a solvability condition that couples the prescribed constrained flow to the geometry through the metric tensor. Using this condition, we construct equilibria in slab, cylindrical and toroidal geometries. In toroidal geometry, the cross-field flow strongly correlates with magnetic-island structure: varying the angular frequency modifies the dominant Fourier harmonic of the radial component of the magnetic field and can drive a transition from a primary ($m=1$) island to secondary ($m=2$) islands. In slab and cylindrical geometries, flow parameters weakly affect island width but strongly modify equilibrium profiles.
\end{abstract}

\section{Introduction}
Kruskal and Kulsrud’s variational model pioneered the construction of three-dimensional (3-D) magnetohydrodynamic (MHD) equilibria \citep{kruskal1958equilibrium}. 
In this model, the toroidal plasma domain is foliated by a continuous set of nested magnetic flux surfaces to which the magnetic field is tangential. The equilibrium configuration is obtained by determining the geometry of these flux surfaces that minimises the magnetic energy functional, subject to the constraints of ideal MHD. It is well known, however, that this model is singular: during minimisation, the frozen-flux constraint of ideal MHD does not allow magnetic islands to form, and in non-axisymmetric geometry the parallel current develops singularities at the position of rational surfaces (\citealp*{rosenbluth1973nonlinear}; \citealp{rodriguez2021islands,huang2023structure}).

To address this problem, relaxed MHD models have been introduced \citep{taylor1974relaxation,taylor1986relaxation,finn1983turbulent,dewar2008relaxed}. In these models, the frozen-in constraint on the magnetic field is relaxed, allowing field lines to undergo topological transitions and enabling the formation of magnetic islands. A particularly successful relaxed model is the multiregion relaxed MHD (MRxMHD), which is implemented numerically in the stepped-pressure equilibrium code (SPEC) (\citealp*{hole2006stepped}; \citealp{hudson2012computation}). In this approach, the equilibrium domain is partitioned into several relaxed regions separated by ideal flux surfaces, with each region assigned a constant pressure. Each region is modelled using Taylor relaxation, in which the local energy functional is minimised while the total magnetic helicity, as well as the poloidal and toroidal fluxes are invariant. At the interfaces between adjacent regions, discontinuities in pressure and magnetic field balance each other so that the total energy density remains continuous. As a result, local Taylor relaxation permits the formation of magnetic islands and chaotic field lines within each region, while the global pressure profile can be prescribed as a desired stepped function. 

The effects of plasma flow on equilibrium properties have been extensively studied in the context of tokamaks and two-dimensional (2-D) equilibria (\citealp{maschke1980exact,hameiri1983equilibrium,Kerner1987ZNaturforschA}; \citealp*{van2000new}; \citealp{guazzotto2004numerical,mcclements2010steady,hole2011identifying}). In contrast, for stellarators, plasma flow is often assumed to be negligible and has therefore received less attention. There is, however, growing evidence that flow in fully 3-D geometries can significantly influence equilibrium structure and magnetic field geometry \citep{boozer1996shielding,simakov2011plasma,cooper2015free,nuhrenberg2021ideal,weitzner2021steady}.

In their original formulations, neither Taylor relaxation nor MRxMHD included plasma flow. \citet{finn1983turbulent} extended single-region Taylor relaxation by incorporating flow through a modification of the underlying variational principle. To enable cross-field flow, they introduced an additional constraint on toroidal angular momentum via a Lagrange multiplier. Because this constraint is physically meaningful only in axisymmetric systems, it was suggested that it be deactivated \emph{ad hoc} in fully 3-D configurations. The Finn--Antonsen variational principle was later extended to MRxMHD in \citet{dennis2014multi} and \citet{qu2020stepped}. In \citet{qu2020stepped}, it was suggested that this model may be applicable to reversed-field-pinch devices, where plasma flow is expected to play an important role. Nevertheless, in the absence of the angular momentum constraint, which is only valid in two dimensions, the resulting flow remains exclusively field-aligned. This is inconsistent with neoclassical theory, which predicts that, in order to avoid viscous damping, plasma flow tends to align with the direction of (quasi-)symmetry rather than the magnetic field \citep{shaing1982neoclassical,hinton1985neoclassical,helander2007rapidO}.

A relaxed MHD model that can naturally accommodate cross-field flow, independent of symmetry assumptions, is therefore required. 
Such a model was developed by \citet{dewar2020time} based on a phase-space Hamiltonian action principle. In this model, which we summarise in \cref{sec:action}, the plasma state is obtained by extremising a phase-space action subject to a set of macroscopic constraints (the magnetic helicity, cross-helicity and entropy), rather than by minimising an energy at fixed helicity as in Taylor relaxation. The resulting phase-space Lagrangian yields two distinct contributions to the flow: a relaxed field-aligned flow, also present in MRxMHD, and a kinematically constrained flow. The latter enables cross-field flow in arbitrary geometries without the introduction of \emph{ad hoc} Lagrange-multiplier constraints of the toroidal angular momentum. In addition, the phase-space formulation allows for the enforcement of cross-helicity conservation and recovers the ideal-MHD equations of motion from the Euler--Lagrange equations. This ensures that the solutions of this model are physically relevant because they satisfy the well-known force-balance equation of the ideal MHD.
 
Despite its potential, the \citet{dewar2020time} theory, which we refer to as RxMHD, remains computationally ambiguous. In particular, it is unclear how the resulting system of Euler–Lagrange equations can be solved in general geometries. This difficulty is exacerbated in the steady-state limit, where the system becomes underdetermined and additional structure is required to obtain unique solutions. In this work, we solve the steady-state equations of the RxMHD in 2-D slab, cylindrical and toroidal geometries. We identify a solvability condition for steady-state solutions of the phase-space relaxed MHD equations. We show that, in the absence of time dependence, the Euler–Lagrange system imposes a non-trivial compatibility condition that couples the prescribed constrained flow to the geometry through the metric tensor. This condition restricts admissible flow profiles and provides a systematic way to construct physically meaningful equilibria in different geometries.

Our numerical results show that the cross-field flow significantly modifies equilibrium profiles in slab and cylindrical geometries. In toroidal geometry, we recover the Finn--Antonsen equations within the new model and demonstrate a strong correlation between plasma flow and magnetic island size. In particular, the angular frequency of flow rotation can substantially alter the width of primary magnetic islands and, in certain limits, can fragment a large primary island into secondary islands. These results demonstrate that plasma flow, especially cross-field flow, can play a decisive role in determining magnetic geometry. Accordingly, we suggest extending the SPEC code to incorporate the more general flow profiles enabled by the RxMHD. Because this is the first numerical investigation of the model, we restrict our analysis to two dimensions. This simplification renders the steady-state problem tractable and allows us to isolate the coupling between cross-field flow and geometry. However, this assumption also means that the magnetic islands obtained here can only have toroidal mode number $n=0$, and a similar investigation of $n\neq 0$ islands is left for future studies.

The rest of the paper is structured as follows. In \cref{sec:action}, we reintroduce the phase-space Lagrangian and its corresponding relaxed-equilibrium problem. In \cref{sec:2Dequi} we introduce an ansatz that limits the solution space of the relaxed equilibrium problem to a class of physical interest and discuss the details of the computational method for the system of equations. In \cref{sec:HKT_cylindrical_slab} we provide an example solution in the Hahm--Kulsrud--Taylor (HKT) slab geometry, followed by the cylindrical geometry. In \cref{sec:toroidal}, an example solution in the toroidal geometry, corresponding to the Finn--Antonsen model, is shown, and the effect of the flow angular frequency on the width of the islands is discussed. We conclude this study in \cref{sec:conclusion}. 

\section{Action principle and Euler--Lagrange equations}\label{sec:action}

The Hamiltonian density of the MHD is
\begin{equation}
    \mathcal{H}^{MHD}[\vb{A}, p, \vb{u}, \rho]=\frac{1}{2}\rho u^2 +\frac{1}{2}B^2 +\frac{p}{\gamma-1},
    \label{eq:GenMHDhamiltonian}
\end{equation}
where $\rho$ is the plasma mass density, $\vb{u}$ is the plasma flow, $\vb{B}$ is the magnetic field, $p$ is the plasma pressure, $\gamma$ is the adiabatic index and $\vb{A}$ is the vector potential defined by $\vb{B}=\curl{\vb{A}}$. The Hamiltonian density is a functional of $\vb{A}$ through the magnetic energy term; we write this term as $\frac{1}{2}B^2$ rather than $\frac{1}{2}\abs{\curl\vb{A}}^2$ for notational convenience. Throughout this work, we adopt units in which the permeability of vacuum is $\mu_0=1$. 
To this Hamiltonian density, we add three `macroscopic' constraints by the method of Lagrange multipliers. The Hamiltonian density of the macroscopic constraints ($H^{MC}$) is
\begin{align}
        \mathcal{H}^{MC}[\vb{A}, p, \vb{u}, \rho,\tau, \mu_B, \nu] =
         -\nu\left(\vb{u} \cdot \vb{B} - \frac{U_0}{V_\Omega}\right) 
        - \mu_B\left(\frac{1}{2} \vb{A} \cdot \vb{B} - \frac{K_0}{V_\Omega}\right) \nonumber \\
           -\tau \left(\frac{\rho}{\gamma - 1} \ln{ \kappa\frac{p}{\rho^\gamma}} - \frac{S_0}{V_\Omega}\right).
        \label{eq:Genhamiltonianf}
\end{align}
Here, $U_0$, $K_0$ and $S_0$ are total cross-helicity, magnetic helicity and entropy, respectively. Here $\nu$, $\mu_B$ and $\tau$ are the corresponding Lagrange multipliers of these three constraints. Here, $V_\Omega$ is the volume of the domain $\Omega$, and $\kappa$ carries the dimensions of $\rho^\gamma/p$ and renders the argument of the logarithm dimensionless.

The phase space Lagrangian is defined by means of the (inverse) Legendre transform, using the Hamiltonians (\cref{eq:GenMHDhamiltonian,eq:Genhamiltonianf}),
\begin{equation}
    \mathcal{L}=\bm{\Pi}\cdot\vb{v}-\mathcal{H}^{MHD}-\mathcal{H}^{MC};\;\;\;\;\;\bm{\Pi}=\rho\vb{u}.
\end{equation}
Here, $\bm{\Pi}=\rho \vb{u}$ is the momentum density. The distinction between the total flow $\vb{u}$ and $\vb{v}$ is intentional and enables the inclusion of the cross-field flow in the model. In this model, $\vb{v}$ is a flow that is constrained by a microscopic kinematic constraint 
\begin{equation}
    \delta \vb{v}= \pdv{\bm{\zeta}}{t}+\vb{v}\cdot\nabla \boldsymbol{\zeta}-\boldsymbol{\zeta}\cdot \nabla\vb{v},
    \label{eq:v_const}
\end{equation}
where $\bm{\zeta}$ is the variation of the Lagrangian coordinates. The total plasma flow, however, is $\vb{u}$; for this reason, in this work we refer to $\vb{u}$ as flow while we refer to $\vb{v}$ as the `constrained' flow. Other than \cref{eq:v_const}, the mass is also microscopically constrained by 
\begin{equation}
    \delta \rho=-\div{\rho\boldsymbol{\zeta}}.
    \label{eq:rho_const}
\end{equation}

The Euler--Lagrange equations are found from Hamilton's action principle \citep{dewar2020time} as 
\begin{subequations}
\begin{align}
p-\tau\rho&=0,\label{eq:isothermal}\\
\vb{u}-\vb{v}-\vb{u}^{Rx}&=0,\label{eq:vu_relation}\\
    \pdv{\vb{u}}{t}+(\curl{\vb{u}})\times\vb{v}+\grad{h_\Omega}&=0,\label{eq:GenBer}\\
\curl{\vb{B}}-\mu_B\vb{B}-\nu\curl{\vb{u}}&=0,\label{eq:GenBelt}
\end{align}
\end{subequations}
where $h_\Omega=\tau \ln (\rho/\rho_\Omega)+(u^2/2)$, and $\vb{u}^{Rx}=\frac{\nu\vb{B}}{\rho}$ is the fully relaxed parallel flow consistent with the cross-helicity constraint. Comparing with the ideal gas law, \cref{eq:isothermal} shows that the Lagrange multiplier $\tau$ is in fact the plasma temperature per unit mass. Also, from \cref{eq:vu_relation}, we can see that the total flow $\vb{u}$ is in fact the summation of the constrained flow $\vb{v}$ and the relaxed field-aligned flow $\vb{u}^{Rx}$. The latter is the flow that we also obtain in the MRxMHD in three dimensions \citep{dennis2014multi,qu2020stepped}. Equation \cref{eq:GenBer} is the equation of motion derived by the explicit variation of Lagrangian coordinates, and \cref{eq:GenBelt} is a generalised Beltrami equation. To obtain a closed system of equations, \cref{eq:isothermal,eq:vu_relation,eq:GenBer,eq:GenBelt} should be amended by the continuity equation:
\begin{equation}
    \pdv{\rho}{t}+\div{\rho\vb{u}}=0.\label{eq:continuity}
\end{equation}

Note that from \cref{eq:vu_relation} we have $\div{\rho \vb{u}}=\div{\rho \vb{v}}=0$. In the time-independent limit, \cref{eq:isothermal,eq:vu_relation,eq:GenBer,eq:GenBelt} read
\begin{subequations}
\begin{align}
p-\tau\rho&=0,\label{eq:isothermal_eq}\\
\vb{u}-\vb{v}-\vb{u}^{Rx}&=0,\label{eq:vu_relation_eq}\\
(\curl{\vb{u}})\times\vb{v}+\grad{h_\Omega}&=0,\label{eq:GenBer_eq}\\
\curl{\vb{B}}-\mu_B\vb{B}-\nu\curl{\vb{u}}&=0,\label{eq:GenBelt_eq}\\
\div{\rho\vb{u}}&=0.\label{eq:continuity_eq}
\end{align}
\label{eq:equilibrium}
\end{subequations}
In these equations, the magnetic field is subject to the boundary condition $\vb{B}\cdot \hat{\vb{n}}=0$ and $\div{\vb{B}}=0$ with $\hat{\vb{n}}$ being the normal to the boundary. The $\div{\vb{B}}=0$ is also implied by \cref{eq:GenBelt_eq}. Equations \cref{eq:isothermal_eq,eq:vu_relation_eq,eq:GenBer_eq,eq:GenBelt_eq} lead to the equilibrium equation of ideal MHD:
\begin{equation}
    \rho\vb{u}\cdot\grad{\vb{u}}=-\grad{p}+\qty(\curl{\vb{B}})\times\vb{B}
    \label{eq:imhd_balance}.
\end{equation}
This relation underscores a key strength of the present model: any solution of this model also satisfies the equations of motion of ideal MHD. We note that the model is not fully equivalent to an ideal MHD equilibrium (by which we mean a steady-state solution of the ideal MHD system of equations), as it does not necessarily enforce the ideal Ohm’s law \citep{dewar2022relaxed}. This essential distinction between the present relaxed model and ideal MHD does not stem from any explicit dissipative term. The variational principle of \citet{dewar2020time} intentionally relaxes the ideal frozen-in constraints, thereby permitting topological changes of the magnetic field lines and the formation of magnetic islands or chaotic field lines that would be forbidden in ideal MHD. The macroscopic constraints (cross-helicity, magnetic helicity and entropy) replace the frozen-in constraints in the variational principle, in the same spirit as Taylor relaxation. The ideal-MHD force-balance equation \cref{eq:imhd_balance}, on the other hand, is fundamental to magnetic confinement and is upheld by the model: it is not imposed as a constraint but emerges from the Euler--Lagrange equations, which is itself a strength of the \citet{dewar2020time} formulation. We also note that, by \cref{eq:vu_relation_eq}, $\vb{u}\times\vb{B}=\vb{v}\times\vb{B}$, so the constraint $\curl(\vb{u}\times\vb{B})=0$ could equivalently be written in terms of $\vb{v}$; we use the total flow $\vb{u}$ to match the conventional ideal-MHD form.

In \citet{dewar2020time}, \cref{eq:isothermal_eq} to \cref{eq:continuity_eq} are referred to as the semirelaxed equilibrium. This terminology distinguishes this state from the fully relaxed equilibrium, which is expected to correspond to the minimum of the Hamiltonian and leads to a fully field-aligned flow. We note that no ideal constraint, such as $\curl(\vb{u} \times \vb{B}) = 0$, is imposed in \cref{eq:isothermal_eq} to \cref{eq:continuity_eq}, as is commonly assumed in equilibrium formulations with flow \citep{Kerner1987ZNaturforschA,guazzotto2004numerical}.




\section{The 2-D equilibrium with purely toroidal constrained flow $\vb{v}$}\label{sec:2Dequi}

\subsection{The coordinate system}
Our coordinate system $\{s,\theta,\zeta\}=\{q^1,q^2,q^3\}$ is defined (inversely) by the position vector $\vb{r}=\vb{r}(s,\theta,\zeta)$. In the curvilinear coordinates, the tangent basis vectors are $\displaystyle{\vb{e}_i=\pdv{\vb{r}}{q^i}}$ and the reciprocal basis vectors are $\vb{e}^i=\grad{q^i}$. For any vector $\vb{w}$, superscripts denote the components in the expansion over the tangent basis, $\vb{w}=w^i\vb{e}_i$ with $w^i=\vb{w}\cdot\vb{e}^i$ (contravariant components), and subscripts denote the components in the expansion over the reciprocal basis, $\vb{w}=w_i\vb{e}^i$ with $w_i=\vb{w}\cdot\vb{e}_i$ (covariant components). The components of the metric tensor are $g_{ij}=\vb{e}_i\cdot\vb{e}_j$, and its inverse has components $g^{ij}=\vb{e}^i\cdot\vb{e}^j$. The Jacobian is therefore $\sqrt{g}=\vb{e}_1\cdot(\vb{e}_2\times \vb{e}_3)=\vb{e}^1\cdot(\vb{e}^2\times \vb{e}^3)$.  In this paper, we assume $\zeta$ denotes the direction of the symmetry; i.e.~all the variable fields are independent of $\zeta$ (the 2-D assumption). Using a general coordinate system, we encode the geometry in the metric tensor. We shall assume that the $\{s,\theta,\zeta\}$ coordinates, as well as the position vector $\vb{r}$, are dimensionless; otherwise, the subscripted and superscripted components might not carry the physical dimension of their corresponding vector quantities \citep{truesdell1953physical}. In this paper we assume $s\in [0,1]$, $\theta \in [0,2\pi)$ and $\zeta \in [0,2\pi)$. We note that these coordinates are not necessarily straight field line coordinates.

To find the position vector (and therefore the coordinate transformations),  we first define the position of the two enclosing boundaries of the domain, labelled by $s=0$ and $ s=1$, as $\vb{\mathcal{R}}_-(s,\theta)$ and $\vb{\mathcal{R}}_+(s,\theta)$. The position vector is then extended from these boundaries to the interior of the domain by linear interpolation; i.e. by defining $\vb{r}=(1-s)\vb{\mathcal{R}}_- +s\vb{\mathcal{R}}_+$.

The magnetic field lines are defined as the solution to the differential equation $\dv{q^i}{\tau_0}=B^i$, where $\tau_0$ is a time-like ordering parameter. The Poincaré map can be obtained by solving the field line equations and collecting the points at which orbits intersect the $\zeta=0$ plane. However, assuming that $B^\zeta$ is never zero, it is easier to integrate the alternative equations
\begin{align}
    \dv{s}{\zeta}&=\frac{B^s}{B^\zeta},\label{eq:poincare_1}\\
    \dv{\theta}{\zeta}&=\frac{B^\theta}{B^\zeta}\label{eq:poincare_2},
\end{align}
to obtain the Poincaré map on $\{s,\theta\}$ plane. The 2-D assumption \(\partial_\zeta = 0\) guarantees that the field lines described by these equations are integrable. The 2-D assumption also
restricts all boundary perturbations to have toroidal mode number \(n = 0\).
Consequently, the only possible magnetic resonance occurs at \(\iota = 0\).
Higher-\(m\) harmonics modify the structure of this resonant layer but do not generate additional island chains. We note that because $\{s,\theta,\zeta\}$ are not straight-field-line coordinates, the rotational transform can be a function of both $s$ and $\theta$. In the following numerical calculations, we mainly report the $\theta$-averaged rotational transform. 

 \subsection{Generalised Beltrami equation}

Equation \cref{eq:GenBelt_eq} is the generalised Beltrami equation. To solve this equation, we start by representing the divergence-free magnetic field as $\vb{B}=\curl{\vb{A}}$ where $\vb{A}$ is the vector potential. Using the gauge freedom, we can assume $\vb{A}=\bar{\vb{A}}+\grad{G}$ so that
\begin{subequations}
\begin{align}
    A_s(s,\theta,\zeta)=\bar{A_s}(s,\theta,\zeta)+\pdv{G}{s}\;(s,\theta,\zeta)&=0,\label{eq:gauge1}\\
    A_\theta(1,\theta,\zeta)=\bar{A_\theta}(1,\theta,\zeta)+\pdv{G}{\theta}\,(1,\theta,\zeta)&=\frac{\psi_\zeta}{2\pi},\label{eq:gauge2}\\
    A_\zeta(1,0,\zeta)=\bar{A_\zeta}(1,0,\zeta)+\pdv{G}{\zeta}\,(1,0,\zeta)&=-\frac{\psi_\theta}{2\pi},\label{eq:gauge3}
\end{align}
\end{subequations}
where $1$ and $0$ are the $s$ coordinates of the two enclosing boundaries, respectively. The quantities $\psi_\theta$ and $\psi_\zeta$ are two constants that are provided as inputs. Equations \cref{eq:gauge2,eq:gauge3} provide boundary conditions for $A_\theta$ and $A_\zeta$ at $s=1$ boundary. In addition, in two dimensions ($\pdv{\zeta}=0$) the $\vb{B}\cdot\hat{\vb{n}}=\vb{B}\cdot \grad{s}$ reads
\begin{equation}
    \pdv{A_\zeta}{\theta} \qty(s=0)=\pdv{A_\zeta}{\theta}  \qty(s=1)=0.
\end{equation}
Therefore, $A_\zeta$ must be constant on both $s=0$ and $s=1$ boundaries. From \cref{eq:gauge3} for the $s=1$ boundary this constant is $-\psi_\theta/(2\pi)$. For the $s=0$ boundary, we use the boundary conditions 
\begin{subequations}
\begin{align}
    \int_0^{2\pi} A_\theta(0,\theta)\dd \theta&=0,\label{eq:left_bou_1}\\
    A_\zeta(0,\theta)&=0.\label{eq:left_bou_2}
\end{align}
\end{subequations}
Using the boundary values of $\vb{A}$, one can show that $\psi_\theta$ and $\psi_\zeta$ are the total enclosed poloidal and toroidal fluxes, respectively; i.e.
\begin{subequations}
\begin{align}
    \int_{0}^{1}\int_0^{2\pi}B^\theta(s,0,\zeta)\sqrt{g}\dd s\dd \zeta&=\int_0^{2\pi} \qty(-A_\zeta(s=1,0)+A_\zeta(s=0,0))\dd \zeta=\psi_\theta,\\
    \int_{0}^{1}\int_0^{2\pi}B^\zeta(s,\theta,0)\sqrt{g}\dd s\dd \theta&=\int_0^{2\pi} \qty(A_\theta(s=1,\theta)-A_\theta(s=0,\theta))\dd \theta=\psi_\zeta.
\end{align}
\end{subequations}

Using the gauge condition \cref{eq:gauge1}, the $s$ and $\theta$ components of \cref{eq:GenBelt_eq} read
\begin{subequations}
\begin{align}
    -\mu_B \pdv{A_\zeta}{\theta}-\nu\pdv{u_\zeta}{\theta}+\pdv{B_\zeta}{\theta}&=0\\
    \mu_B \pdv{A_\zeta}{s}+\nu\pdv{u_\zeta}{s}-\pdv{B_\zeta}{s}&=0
\end{align}
\end{subequations}
where $B_\zeta=g_{31}B^s+g_{32}B^\theta+g_{33}B^\zeta$. These two equations can be combined in 
\begin{equation}
    \mu_BA_\zeta+\nu u_\zeta-B_\zeta=\mathcal{C},
    \label{eq:belt_12}
\end{equation}
where $\mathcal{C}$ is a constant determined by the boundary conditions. Using \cref{eq:vu_relation_eq}, \cref{eq:belt_12} can be written as
\begin{equation}
    \mu_BA_\zeta+\nu v_\zeta+\qty[(M^{Rx}_\parallel)^2-1] B_\zeta=\mathcal{C},
    \label{eq:belt_12_cart}
\end{equation}
where $M^{Rx}_\parallel$ is the parallel Alfv{\'e}n Mach number defined by
\begin{equation}
\qty(M^{Rx}_\parallel)^2\equiv \frac{(u^{Rx})^2}{c_A^2}=\frac{\nu ^2}{\rho},
\label{eq:MachN}
\end{equation}
and $c_A\equiv ({B^2/\rho})^{1/2}$ is the Alfv{\'e}n speed. 
Equation \cref{eq:belt_12} should be solved in conjunction with the $\zeta$ component of \cref{eq:GenBelt_eq}:
\begin{equation}
    -\mu_B B^\zeta-\frac{\nu}{\sqrt{g}} \qty(\pdv{u_\theta}{s}-\pdv{u_s}{\theta})+\frac{1}{\sqrt{g}} \qty(\pdv{B_\theta}{s}-\pdv{B_s}{\theta})=0
    \label{eq:belt_3}
\end{equation}

\subsection{The equation of motion}\label{subsec:eqofM}

Neoclassical considerations motivate the study of flow that is approximately tangent to contours of $B$, i.e.\ $\vb{u}\cdot\nabla B \approx 0$ \citep{shaing1982neoclassical,hinton1985neoclassical,helander2007rapidO}. In 2-D equilibria
where $B$ varies predominantly across $(s,\theta)$, this suggests that the dominant
component of the flow lies along the symmetry direction ($\zeta$).
 To achieve this, we prescribe $\vb{v}$ to be 
\begin{equation}
    \vb{v}=v^\zeta\vb{e}_{\zeta},
    \label{eq:v_troidal}
\end{equation}
so that the only non-zero component of $\vb{v}$ is $v^\zeta$. In two dimensions, this assumption also implies $\div{\rho\vb{u}}=\displaystyle{\pdv{(\sqrt{g}\rho v^\zeta)}{\zeta}}=0$; i.e.~the continuity equation \cref{eq:continuity_eq} is satisfied. Using \cref{eq:vu_relation_eq}, $\vb{u}$ can be decomposed into the  poloidal and toroidal components
\begin{equation}
    \vb{u}=\frac{\nu}{\rho}\qty(B^s\vb{e}_s+B^\theta\vb{e}_\theta)+\qty(\frac{\nu B^\zeta}{\rho}+v^\zeta)\vb{e}_\zeta
    \label{eq:pol_toroid_flow}
\end{equation}
Therefore, like the magnetic field, the total flow $\vb{u}$ is generally helical. To be consistent with the expectations of neoclassical transport theory, one can assume  $\nu\ll 1$ and $\rho \sim O(1)$ (hence $M^{Rx}_\parallel\ll 1$) so that the poloidal flow as defined by \cref{eq:pol_toroid_flow} is much smaller than $v^\zeta$.

Another useful decomposition of $\vb{u}$ is into its field-aligned and cross-field components
\begin{equation}
    \vb{u}=\vb{u}_\parallel+\vb{u}_\perp= \frac{\vb{u}\cdot \vb{B}}{B^2}\vb{B}+ \frac{1}{B^2}\vb{B}\times \qty( \vb{u}\times \vb{B}).
\end{equation}
Using \cref{eq:vu_relation_eq,eq:v_troidal}
\begin{subequations}
   \begin{align}
    \vb{u}_\parallel&=\qty(\frac{\nu}{\rho}+\frac{B_\zeta v^\zeta}{B^2})\vb{B},\\
    \vb{u}_\perp&=v^\zeta\vb{e}_\zeta-\frac{B_\zeta v^\zeta}{B^2}\vb{B}, 
\end{align} 
\end{subequations}
and the parallel and perpendicular components of the flow are
\begin{subequations}
\begin{align}
    u_\parallel&=\frac{\nu B^2+\rho v^\zeta B_\zeta}{\rho B},\label{eq:parallel_flow}\\
    u_\perp&=\frac{\abs{v^\zeta}}{B}\sqrt{g_{33}B^2-B_\zeta^2}=\frac{\abs{v^\zeta}}{B}\sqrt{g_{33}B_p^2-g_{31}B^sB_\zeta-g_{32}B^\theta B_\zeta},\label{eq:perp_flow}
\end{align}
\end{subequations}
where $B_p\equiv \sqrt{B_sB^s+B_\theta B^\theta}$ is the norm of poloidal field. We define the `cross-field flow ratio' as $(u_\perp/u_\parallel)$. 

Because the O point of a magnetic island is a fixed point of the Poincaré map, 
\cref{eq:poincare_1,eq:poincare_2} imply that $B^s = B^\theta = 0$ at the O point. 
Consequently, \cref{eq:perp_flow} predicts that the cross-field component of the flow 
vanishes locally and the flow becomes field-aligned at the island centre.
This prediction is consistent with experimental observations in tokamaks and 
stellarators, which show that cross-field flow vanishes or is strongly suppressed 
at the O point of low-level magnetic islands 
\citep{estrada2016plasma,jiang2018influence}.
Importantly, this behaviour is not 
imposed in the present RxMHD formulation but arises self-consistently from the 
variational structure of the model together with the constrained-flow ansatz 
\cref{eq:v_troidal}. This provides additional justification for the physical 
consistency of the ansatz adopted here.

Using \cref{eq:v_troidal} one can show that $(\curl{\vb{u}})\times\vb{v}=-v^\zeta\grad{u_\zeta}$ and therefore \cref{eq:GenBer_eq} reads
\begin{equation}
v^\zeta\grad{u_\zeta}=\grad{h_p}+\frac{\grad{\qty(u^{\zeta}u_{\zeta})}}{2},
\label{eq:ber_grad_form}
\end{equation}
where $h_p\equiv ((u^s u_s+u_\theta u^\theta)/2)+\tau\ln(\rho/\rho_\Omega)$ is the poloidal part of the $h$. Taking the curl of \cref{eq:ber_grad_form}, and using \cref{eq:vu_relation_eq} we get
\begin{equation}
\grad{v^\zeta}\times \grad{u_\zeta^{Rx}}+v^\zeta\grad{v^\zeta}\times\grad{g_{33}}=0.\label{eq:toroid_cond}
\end{equation}
Equation~\cref{eq:toroid_cond} constitutes a \emph{solvability condition} for the steady-state semirelaxed equilibrium equations. In the absence of time dependence, the system~\cref{eq:equilibrium} is generically underdetermined, and not all prescribed constrained flows $\vb{v}$ lead to admissible equilibria. The solvability condition expresses a necessary compatibility between the constrained flow $v^\zeta$, the relaxed field-aligned flow $u^{Rx}_\zeta$, and the geometry through the metric coefficient $g_{33}$.
This result plays a central role in what follows: it dictates which forms of $v^\zeta$ are admissible in a given geometry and explains why flow parameters affect equilibrium structure differently in slab, cylindrical and toroidal configurations.

An alternative form of \cref{eq:toroid_cond} can be calculated by using \cref{eq:vu_relation_eq,eq:belt_12} to obtain
\begin{equation}
    \grad{u_\zeta^{Rx}}=-\frac{1}{\nu}\grad{(\mu_BA_\zeta-B_\zeta+\nu v_\zeta)}.
\end{equation}
Using this relation, \cref{eq:toroid_cond} reads
\begin{equation}
    \grad{(-\mu_BA_\zeta+B_\zeta)}\times \grad{v^\zeta}=0.
    \label{eq:toroid_cond_simple}
\end{equation}
To satisfy this equation, it is sufficient that $v^\zeta$ is an arbitrary function of $-\mu_BA_\zeta+B_\zeta$.
Using \cref{eq:v_troidal,eq:belt_12}, one can reduce \cref{eq:belt_3} to
\begin{gather}
  -\mu_Bg^{31}B_s-\mu_Bg^{32}B_\theta -\mu_Bg^{33}\qty(\frac{\mathcal{C}-\mu_BA_\zeta-\nu v^\zeta g_{33}}{\nu^2/\rho-1})+\frac{1}{\sqrt{g}}\pdv{s}\qty[\frac{\qty(\nu^2/\rho-1)}{\sqrt{g}}\pdv{A_\zeta}{s}]\nonumber\\
  +\frac{1}{\sqrt{g}}\pdv{\theta}\qty[\frac{\qty(\nu^2/\rho-1)}{\sqrt{g}}\pdv{A_\zeta}{\theta}]-\frac{\nu}{\sqrt{g}}\qty(\pdv{(g_{32}v^\zeta)}{s}-\pdv{(g_{31}v^\zeta)}{\theta})=0.
    \label{eq:belt_Azeta}
\end{gather}

\subsection{The numerical methods and convergence plots}
In each geometry, \cref{eq:belt_12,eq:belt_3,eq:ber_grad_form} are solved to obtain an equilibrium solution. As shown later, \cref{eq:toroid_cond} allows \cref{eq:ber_grad_form} to be reduced to an algebraic equation prior to the numerical analysis. Numerical solutions are obtained using Dedalus v3
 \citep{2020PhRvR...2b3068B}, a Python-based framework that enables symbolic specification of partial differential equations. The underlying numerical solver employs tau-spectral methods, which are widely used in magnetic confinement equilibrium studies.

\begin{figure}
    \centering
    \begin{overpic}[width=0.49\linewidth]{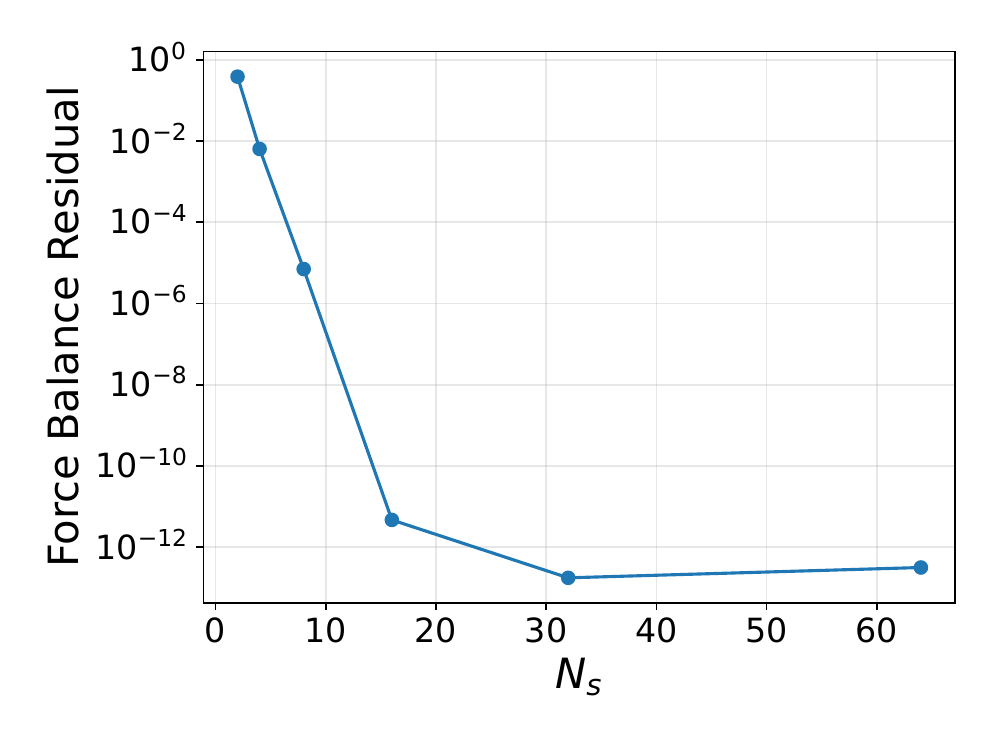}
        \put(10,70){\textbf{(a)}}
        \phantomsubcaption\label{fig:conv_s}
    \end{overpic}
    \hfill
    \begin{overpic}[width=0.49\linewidth]{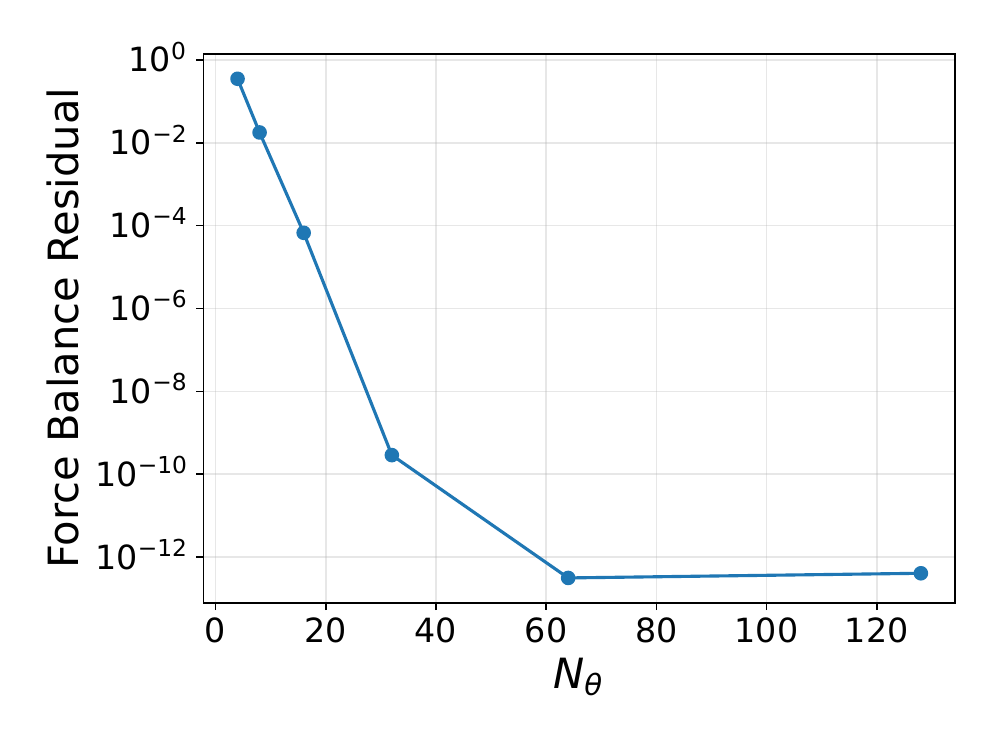}
        \put(10,70){\textbf{(b)}}
        \phantomsubcaption\label{fig:conv_th}
    \end{overpic}
    \caption{Numerical convergence of the Dedalus solver in the toroidal geometry. (a) Force-balance residual versus the Chebyshev resolution $N_s$, with $N_\theta=64$ fixed. (b) Force-balance residual versus the Fourier resolution $N_\theta$, with $N_s=64$ fixed.}
    \label{fig:force_balance_res_vs_M}
\end{figure}

For the 2-D studies presented here, we use Chebyshev expansions in the $s$ direction and Fourier expansions in the $\theta$ direction. To assess the numerical accuracy of our computations, we perform a series of convergence tests. We denote by $N_s$ the number of Chebyshev polynomials used in the $s$ direction and by $N_\theta$ the number of Fourier modes used in the $\theta$ direction. Figure~\ref{fig:force_balance_res_vs_M} shows the residue of the ideal MHD force-balance equation \cref{eq:imhd_balance} as a function of $N_s$ and $N_\theta$ separately: in \cref{fig:conv_s} $N_s$ is varied with $N_\theta=64$ fixed, and in \cref{fig:conv_th} $N_\theta$ is varied with $N_s=64$ fixed. In each case the resolution that is held fixed is large enough that it does not limit the residual, so that each panel isolates the convergence in a single direction. In both directions, the residual initially decreases exponentially over the first four refinement steps, as expected for Chebyshev--Fourier (pseudo-)spectral methods applied to smooth functions \citep{boyd2001chebyshev}. Beyond this, the decrease plateaus at a floor of order $10^{-12}$, where the finite solver tolerance and machine round-off error become the dominant sources of numerical error. The convergence test here is reported only in the toroidal geometry corresponding to the example solution of \cref{sec:toroidal}. We have, however, observed similar convergence in all numerical solutions reported in this study.

Although we did not need to use this capability in this study, Dedalus v3 is also a parallel framework designed for high-performance computing. However, it has some limitations that may make it less suitable for large-scale magnetic confinement projects. In particular, it does not support custom coordinate systems of the type used in this work. As a result, we were required to express the governing equations explicitly in terms of scalar partial derivatives and to handle metric components manually. Another limitation of Dedalus v3 is that it relies on direct matrix inversion to solve the linear systems arising from spectral discretisation and Newton iterations, rather than more modern iterative solvers such as GMRES \citep{saad2003iterative}. Consequently, it does not provide any solution for an underdetermined system of equations, as such systems will lead to a rank-deficient matrix that cannot be inverted. 

\section{Example solutions in the HKT slab geometry and the cylindrical geometry}\label{sec:HKT_cylindrical_slab}
\subsection{The HKT slab geometry}
\begin{figure}
    \centering

    \begin{overpic}[width=0.49\linewidth]{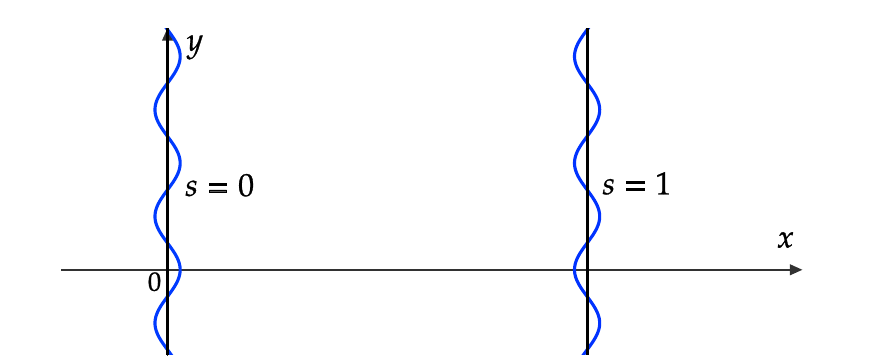}
        \put(10,70){\textbf{(a)}}
    \end{overpic}
    \hfill
    \begin{overpic}[width=0.49\linewidth]{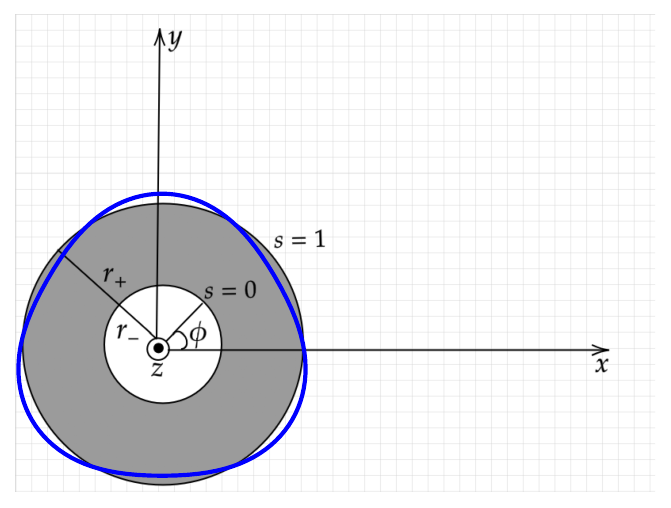}
        \put(10,70){\textbf{(b)}}
    \end{overpic}

    \caption{(a) The HKT slab geometry. (b) Cylindrical geometry. The perturbed boundaries are shown in blue.}
    \label{fig:HKT_Slab}
\end{figure}

In the Cartesian coordinate system, the position vector is $\vb{r}=x\hat{x}+y\hat{y}+z\hat{z}$, where $\hat{x}$, $\hat{y}$ and $\hat{z}$ are the usual unit vectors of Cartesian coordinates. The domain of our HKT slab is shown in \cref{fig:HKT_Slab}. The left-hand boundary and right-hand boundaries, are positioned at $\vb{\mathcal{R}}_-=\delta\cos{3\theta}\hat{x}+\theta\hat{y}+\zeta \hat{z}$ and $\vb{\mathcal{R}}_+=(1-\delta \cos{3\theta})\hat{x}+\theta\hat{y}+\zeta \hat{z}$ , respectively. Using linear interpolation, the position vector in the whole domain is $\vb{r}=\qty[(1-s)\delta\cos{3\theta}+s(1-\delta\cos{3\theta})]\hat{x}+\theta\hat{y}+\zeta \hat{z}$.  Therefore, transformation from the $\{x,y, z\}$ coordinates to the curvilinear coordinates $\{s,\theta,\zeta\}$ is
\begin{align}
    x&=(1-s)\delta\cos(3\theta)+s(1-\delta\cos{3\theta}),\\
    y&=\theta,\\
    z&=\zeta.
\end{align}
In the HKT slab geometry, the metric components are $g_{ij}=\pdv{x}{q^i}\pdv{x}{q^j}+\pdv{y}{q^i}\pdv{y}{q^j}+\pdv{z}{q^i}\pdv{z}{q^j}$ which gives
\begin{equation}
g_{ij}=\left(
\begin{array}{ccc}
 g_{11} & g_{12} & 0 \\
g_{21}& g_{22} & 0 \\
 0 & 0 & 1 \\
\end{array}
\right),
\end{equation}
where $g_{11}=(1-2 \delta  \cos (3 \theta ))^2$, $g_{12}=g_{21}=3 \delta  (2 s-1) (\sin (3 \theta )-\delta  \sin (6 \theta ))$ and $g_{22}=9 \delta ^2 (1-2 s)^2 \sin ^2(3 \theta )+1$. Therefore, we have
\begin{equation}
    g_{33}=1,
    \label{eq:cart_g_33}
\end{equation}
and \cref{eq:toroid_cond} reads
\begin{equation}
    \grad{u_\zeta^{Rx}}\times\grad{v^\zeta}=0.
    \label{eq:uRx_v_cond_grad}
\end{equation}
In the neighbourhood of points where $\pdv{u^{Rx}_\zeta}{s}\neq 0$, one can invert the function $u_\zeta=u_\zeta^{Rx}(s,\theta)$ to $s=s(u^{Rx}_\zeta,\theta)$ and write $v^\zeta = v^\zeta(u^{Rx}_\zeta,\theta)$, which upon using \cref{eq:uRx_v_cond_grad}, implies $v^\zeta$ is a function of only $u_\zeta^{Rx}$; i.e.
\begin{equation}
   v^\zeta=\mathcal{F}(\nu u^{Rx}_\zeta)=\mathcal{F}(\qty(M^{Rx}_\parallel)^2B_\zeta)
   \label{eq:uRx_v_cond_fun}
\end{equation} 
for an arbitrary function $\mathcal{F}$. Note that \cref{eq:uRx_v_cond_fun} alone is sufficient for \cref{eq:uRx_v_cond_grad} to hold. Also, from \cref{eq:vu_relation_eq}, \cref{eq:uRx_v_cond_fun} is equivalent to $u^\zeta=\mathcal{F}\qty(u^{Rx}_\zeta)$. Although $\mathcal{F}$ is an arbitrary function, one can argue that in its simplest form, it should be a linear function.  Assuming the function $\mathcal{F}$ is analytical in a neighbourhood of $\qty(M^{Rx}_\parallel)^2B_\zeta=0$, it can be expanded as 
\begin{equation}
    \mathcal{F}(\qty(M^{Rx}_\parallel)^2B_\zeta)=\sum_{n=0}^\infty c_nB_\zeta^n\qty(M^{Rx}_\parallel)^{2n}.
    \label{eq:F_expansion}
\end{equation}

 If $M_\parallel^{Rx}\ll 1$, in \cref{eq:F_expansion} one can ignore the $(M_\parallel^{Rx})^m$ term for $m\geq 4$; i.e. 
\begin{equation}
   v^\zeta=\alpha u^{Rx}_\zeta + \omega,
    \label{eq:v_z_lin}
\end{equation}
where $c_0 \nu=\omega$, and $c_1 \nu=\alpha$. 
Equation \cref{eq:v_z_lin} also means 
\begin{equation}
    u^\zeta=(\alpha+1) u^{Rx}_\zeta+\omega.
    \label{eq:u_z_lin}
\end{equation}
Physically, $\omega$ is like the angular frequency of plasma rotating around the axis of symmetry, while $(\alpha+1) u^{Rx}_\zeta$ is a small perturbation which is taken as a function of $u^{Rx}_\zeta$ to satisfy \cref{eq:uRx_v_cond_grad}. Using this equation, \cref{eq:belt_12} reads
\begin{equation}
    \mu_BA_\zeta+\qty[(\alpha+1) \qty(M^{Rx}_\parallel)^2-1] B_\zeta=\mathcal{C},
    \label{eq:belt_12_cart_lin}
\end{equation}
where $\omega$ is absorbed in the $\mathcal{C}$. This equation has an obvious singularity in $\qty(M^{Rx}_\parallel)^2=1/(\alpha+1)$, or equivalently $(\alpha+1)\nu^2=\rho$. Because we are assuming $\qty(M^{Rx}_\parallel)^2\ll 1$, this singularity is avoided as long as $\alpha$ is not too large. Nevertheless, we require a high numerical resolution to resolve the equilibrium points near this singularity accurately.

Using \cref{eq:v_z_lin,eq:cart_g_33}, \cref{eq:ber_grad_form} reads
\begin{equation}
    \frac{u^{Rx,s} u^{Rx}_s+u_\theta^{Rx} u^{Rx,\theta}}{2}+\tau\ln{\frac{\rho}{\rho_\Omega}}+(\alpha+1)\frac{(u^{Rx}_\zeta)^2}{2}=0,
    \label{eq:ber_grad_form_cart}
\end{equation}
where any constant on the right-hand side is absorbed in the arbitrary constant $\rho_\Omega$.

We solve \cref{eq:ber_grad_form_cart,eq:belt_12_cart_lin,eq:belt_3} for $A_\theta$, $A_\zeta$ and $\rho$, subject to the boundary conditions of \cref{eq:gauge2}--\cref{eq:left_bou_2}. Other quantities in \cref{eq:equilibrium} can then be obtained from these three quantities.  Although this reduced system of equations does not depend on $\omega$, the flow will depend on it. Using \cref{eq:parallel_flow,eq:perp_flow} the cross-field flow ratio is calculated as
\begin{equation}
    \frac{u_\perp}{u_\parallel}=\frac{\abs{\nu\alpha B_\zeta B_p+\rho \omega B_p}}{\nu B^2+\nu \alpha B_\zeta^2+\rho \omega B_\zeta}.
    \label{eq:flow_anis_slab}
\end{equation}

\begin{figure}
\centering

\begin{overpic}[width=0.49\textwidth]{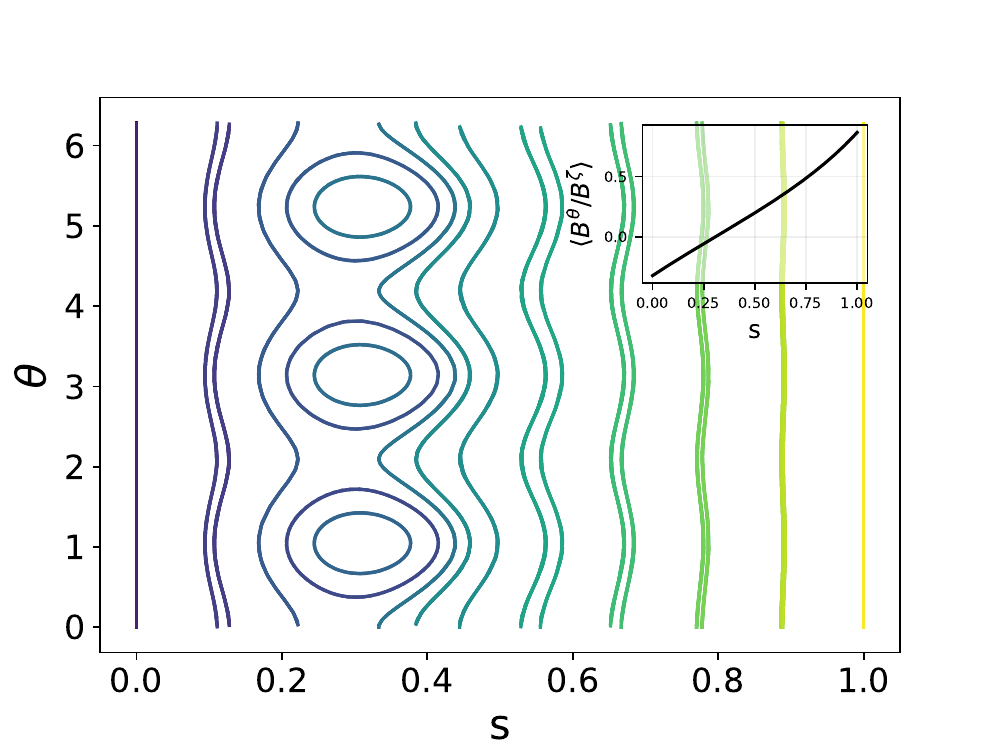}
    \put(10,70){\textbf{(a)}}
    \phantomsubcaption\label{fig:slab_poinc}
\end{overpic}
\hfill
\begin{overpic}[width=0.49\textwidth]{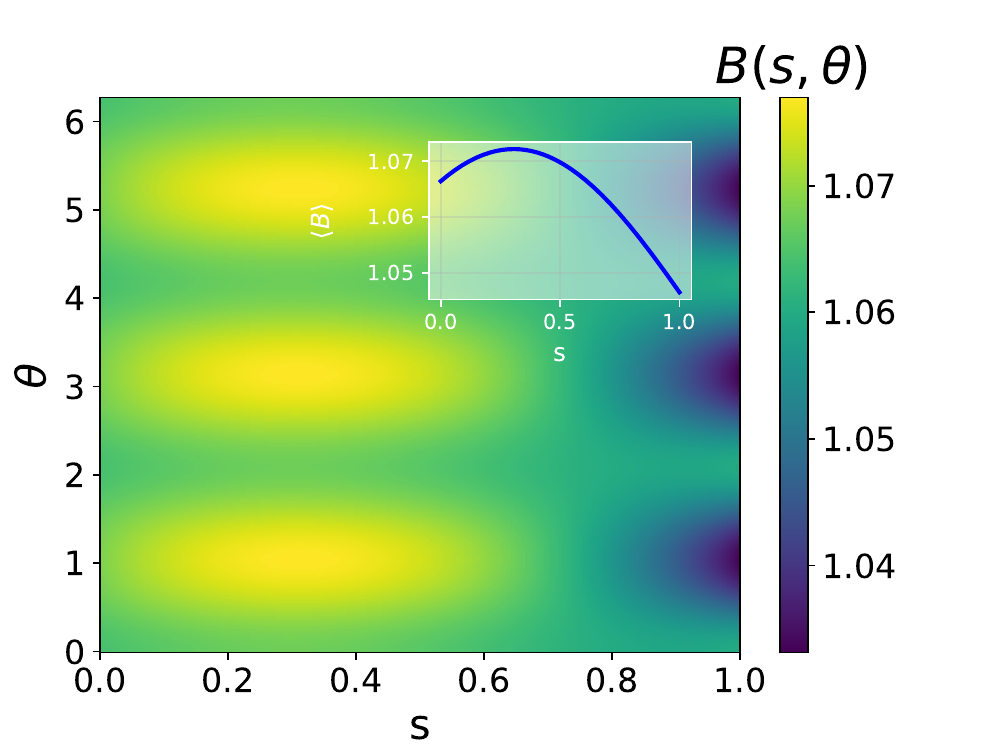}
    \put(10,70){\textbf{(b)}}
    \phantomsubcaption\label{fig:slab_B_norm}
\end{overpic}

\vspace{2mm}

\begin{overpic}[width=0.49\textwidth]{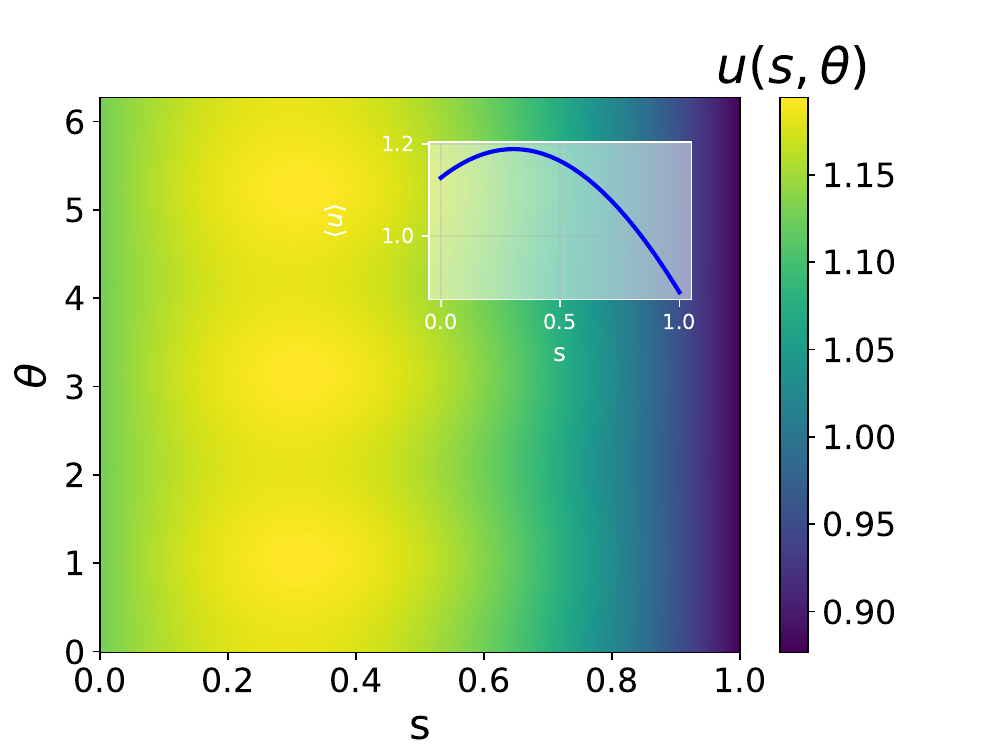}
    \put(10,70){\textbf{(c)}}
    \phantomsubcaption\label{fig:slab_u_norm}
\end{overpic}
\hfill
\begin{overpic}[width=0.49\textwidth]{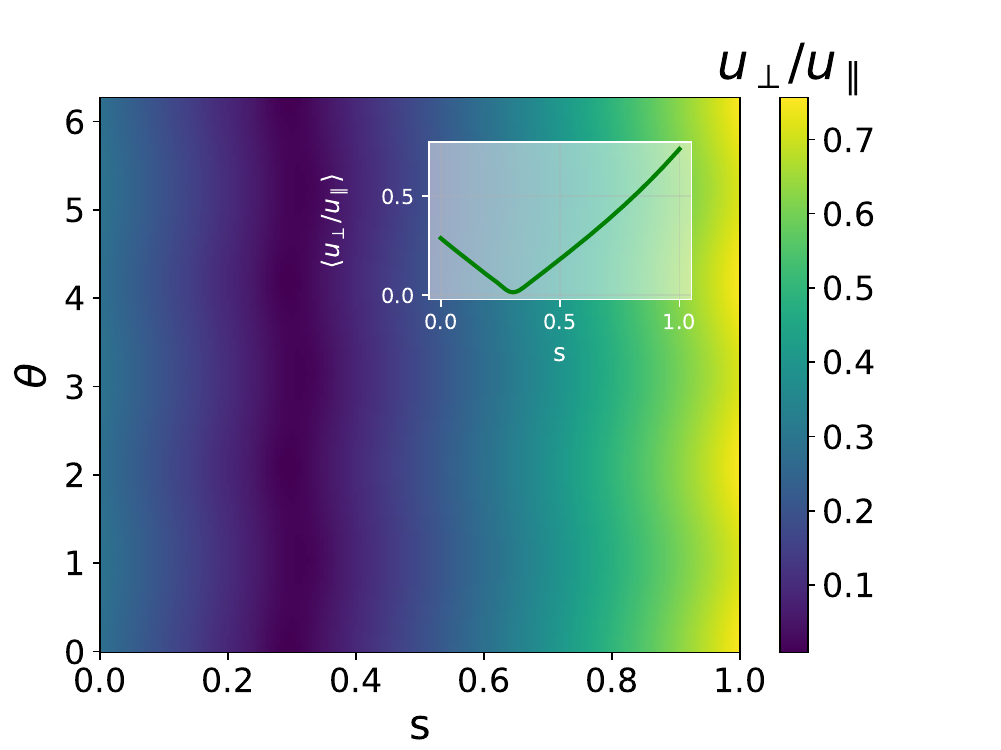}
    \put(10,70){\textbf{(d)}}
    \phantomsubcaption\label{fig:slab_u_perp_parallel}
\end{overpic}

\caption{Example solutions in an HKT slab geometry. (a) Poincaré plot of the magnetic field, (b) magnetic field norm, (c) flow norm and (d) cross-field flow ratio. Here $\delta=0.02$ and $\alpha=10$ used along with the parameters of \cref{tab:parameters}.}
\label{fig:slab}
\end{figure}

Figure~\ref{fig:slab}(\textit{a--d}) shows a solution in the HKT slab geometry using the parameters listed in \cref{tab:parameters}. 
In these parameters the $\nu$ is chosen as a relatively small number so that, with $\rho$ of order unity in our dimensionless units, $M^{Rx}_\parallel\ll 1$ as we discussed in \cref{subsec:eqofM}. On the other hand, $\tau$ is chosen as a relatively large dimensionless value, corresponding to a hot, low-collisionality plasma in our units, while keeping all quantities of order unity. The magnetic fluxes $\psi_\theta$ and $\psi_\zeta$ are tuned so that the rotational transform profile passes through $\iota=0$ and an island is formed in the domain. This, therefore, allows us to study the interaction of magnetic islands with flow.
As we see in \cref{fig:slab_poinc}, at about $s=0.3$ the rotational transform passes through $\iota=0$ and therefore the resonant islands appear in the Poincaré plot. In this paper, the $<>$ sign always denotes the $\theta$-averaged values. The $\theta$-averaged norm of the magnetic field and flow is maximum at the island position.  The cross-field flow vanishes at $s\approx 0.3$, so that the flow becomes field-aligned there, which corresponds to the position of the O points of the island. As mentioned, this is expected from \cref{eq:perp_flow} regardless of the geometry.
\begin{table}
    \centering
    \begin{tabular}{|c|c|}
    \hline
    Parameter & Value \\ \hline
    $\mu_B$ & 1 \\ \hline
    $\nu$ & 0.1 \\ \hline
    $\tau$ & 8 \\ \hline
    $\psi_\theta$ & 0.2 \\ \hline
    $\psi_\zeta$ & 1 \\ \hline
    $\rho_\Omega$ & 1 \\ \hline
    \end{tabular}
    \caption{Parameters used in numerical experiments.}
    \label{tab:parameters}
\end{table}

We emphasise that all quantities in this work are treated as dimensionless, consistent with the dimensionless coordinates introduced in \cref{sec:2Dequi} and the choice $\mu_0=1$. Equivalently, all variables and parameters are expressed in a single self-consistent system of units in which a reference length, magnetic field and mass density are set to unity. The Lagrange multipliers $\mu_B$, $\nu$ and $\tau$ are therefore likewise dimensionless in these units. The values in \cref{tab:parameters} are not intended to model a specific device; rather, they are chosen so that the equilibrium lies in the regime of interest for this study, namely a small parallel Alfvén Mach number, $(M^{Rx}_\parallel)^2=\nu^2/\rho\ll 1$ (with $\rho$ of order unity in these units), together with magnetic fluxes that place a resonant surface within the domain so that an $\iota=0$ island forms. The purpose of the numerical examples is to demonstrate that the RxMHD steady-state problem is well-defined and to expose the coupling between cross-field flow and geometry, rather than to reproduce the parameters of a particular experiment. We note that, because all quantities are of order unity in these units, the corresponding local plasma beta $\beta = 2p/B^2$ is also of order unity in the numerical examples. This is a consequence of the order-unity parameter choice adopted for numerical convenience and is not a prediction of the model: a low-$\beta$ regime more representative of tokamaks and stellarators can be accessed by choosing $p \ll B^2$, for example by choosing a smaller $\tau$ or larger magnetic fluxes.

\subsection{The cylindrical geometry} 
We next consider a hollow cylinder geometry shown in \cref{fig:HKT_Slab}. In the cylindrical geometry, the position vector is $\vb{r}=R\hat{r}(\phi)+Z\hat{z}$, where $\hat{r}$ and $\hat{z}$ are the unit vector basis of the cylindrical coordinate. The inner boundary and outer boundaries, are positioned at $\vb{\mathcal{R}}_-=r_-\hat{r}+\zeta\hat{z}$ and $\vb{\mathcal{R}}_+=(r_+-\delta \cos{3\theta})\hat{r}+\zeta\hat{z}$ , respectively. Using these relations, the transformation from the $\{R,\phi, Z\}$ coordinates to the curvilinear coordinates $\{s,\theta,\zeta\}$ is
\begin{align}
    R&=(1-s)r_-+s[r_+-\delta\cos(3\theta)],\\
    \phi&=\theta,\\
    z&=\zeta,
\end{align}
where $r_-$ and $r_+$ are the inner and outer radii of the hollow cylinder. In the cylindrical geometry, the metric components are $g_{ij}=\pdv{R}{q^i}\pdv{R}{q^j}+\pdv{\hat{r}}{q^i}\cdot\pdv{\hat{r}}{q^j}R^2+\pdv{z}{q^i}\pdv{z}{q^j}$, which gives
\begin{equation}
g_{ij}=\left(
\begin{array}{ccc}
 g_{11}
 &
g_{12}
 & 0
 \\[6pt]
g_{21}
 &
g_{22}
 & 0
 \\[6pt]
 0 & 0 & 1
\end{array}
\right),
\end{equation}
where $g_{11}=(\delta\cos(3\theta)+r_- - r_+)^{2}$ $g_{12}=g_{21}= -3\,\delta\, s\, \sin(\theta)\,\bigl(2\cos(2\theta)+1\bigr) \bigl(\delta\cos(3\theta)+r_- - r_+\bigr)$ and $g_{22}=   \bigl(r_- (s-1)-r_+ s\bigr)\Bigl(r_- (s-1)-r_+ s+2\delta s \cos(3\theta)\Bigr)+\delta^{2} s^{2}\bigl(5-4\cos(6\theta)\bigr)$.  Because in the cylindrical geometry we have $g_{33}=1$, the condition of \cref{eq:toroid_cond} is satisfied in the same way that it was in the HKT slab; i.e. by defining $v_\zeta$ in the form of \cref{eq:v_z_lin}. Therefore, the system of equations is the same system that was solved for the HKT slab. 

\begin{figure}
\centering

\begin{overpic}[width=0.49\textwidth]{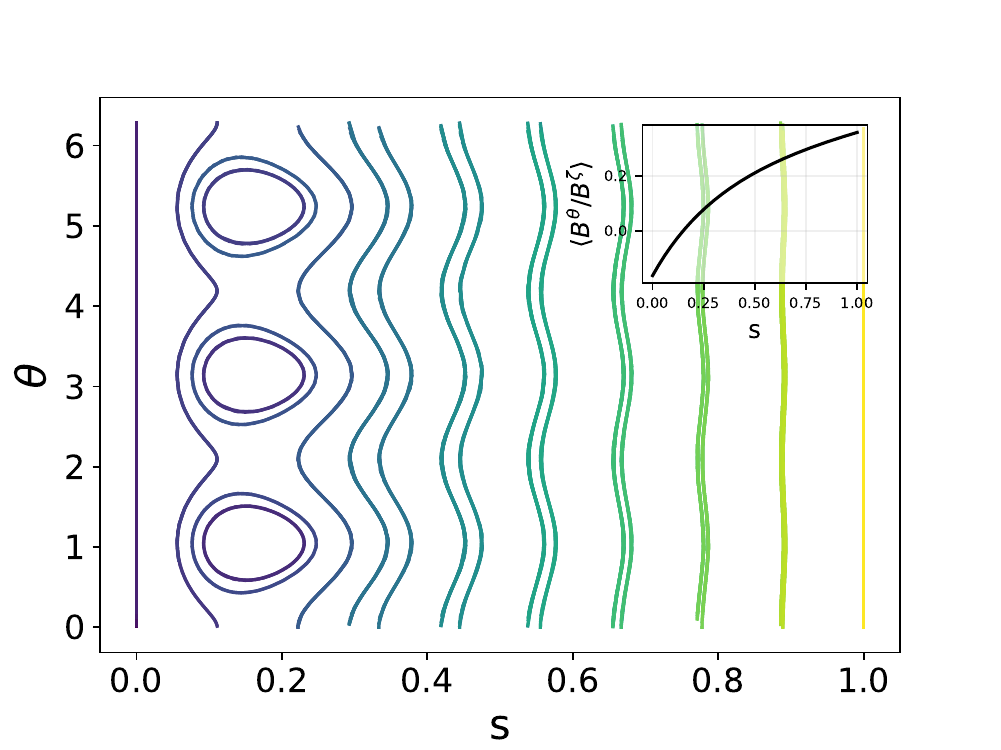}
    \put(10,70){\textbf{(a)}}
    \phantomsubcaption\label{fig:cyl_poinc}
\end{overpic}
\hfill
\begin{overpic}[width=0.49\textwidth]{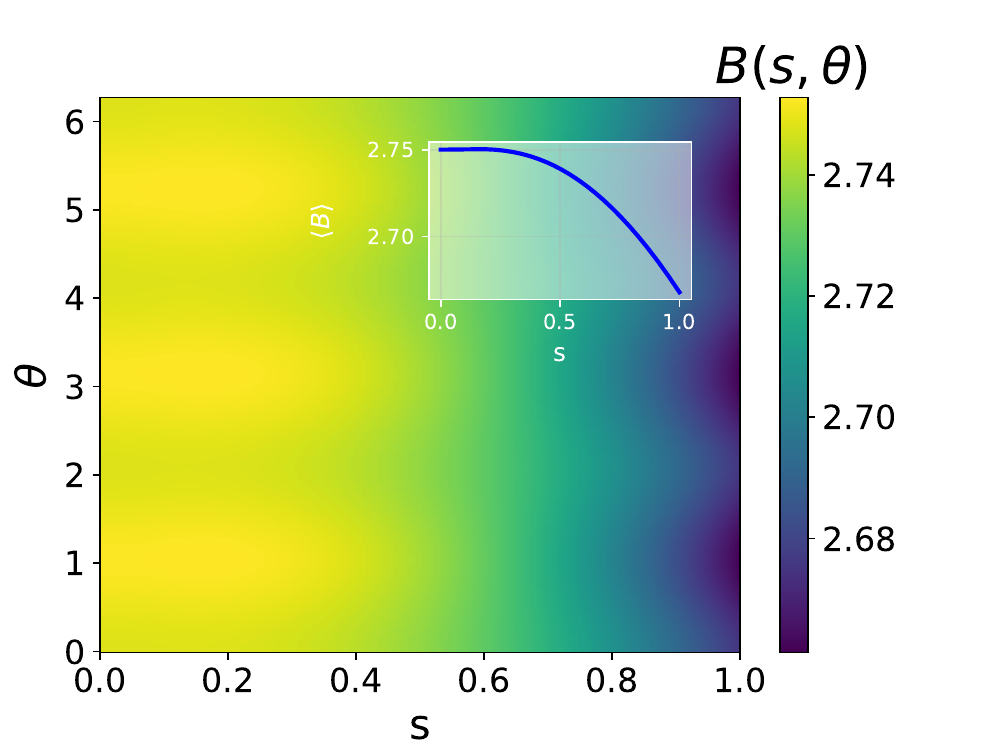}
    \put(10,70){\textbf{(b)}}
    \phantomsubcaption\label{fig:cyl_B_norm}
\end{overpic}

\vspace{2mm}

\begin{overpic}[width=0.49\textwidth]{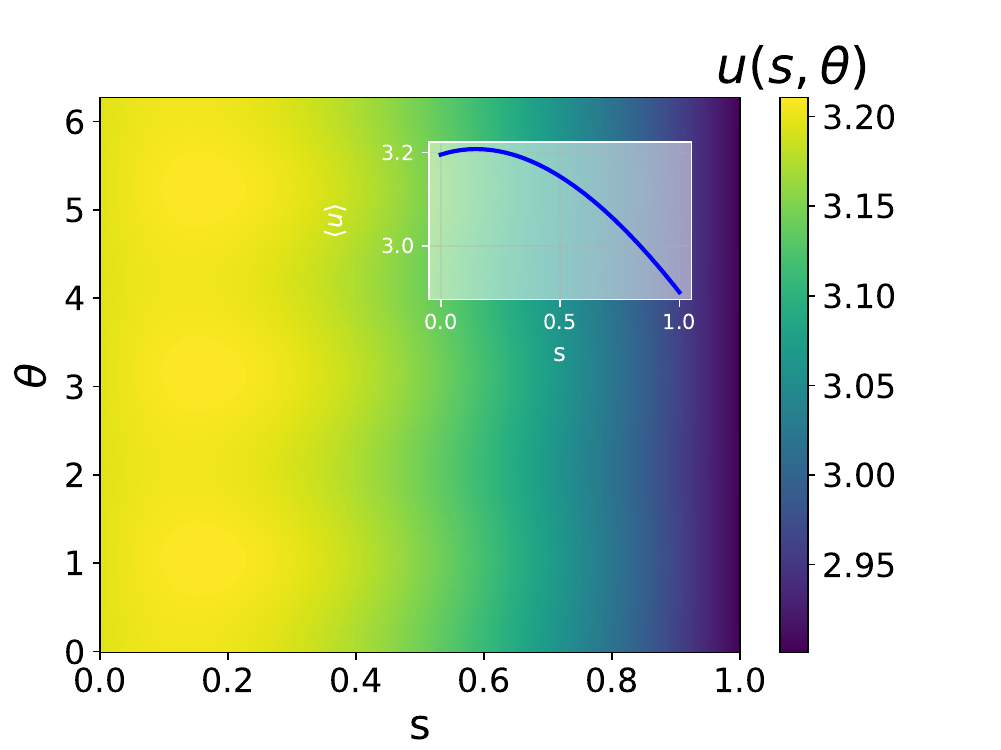}
    \put(10,70){\textbf{(c)}}
    \phantomsubcaption\label{fig:cyl_u_norm}
\end{overpic}
\hfill
\begin{overpic}[width=0.49\textwidth]{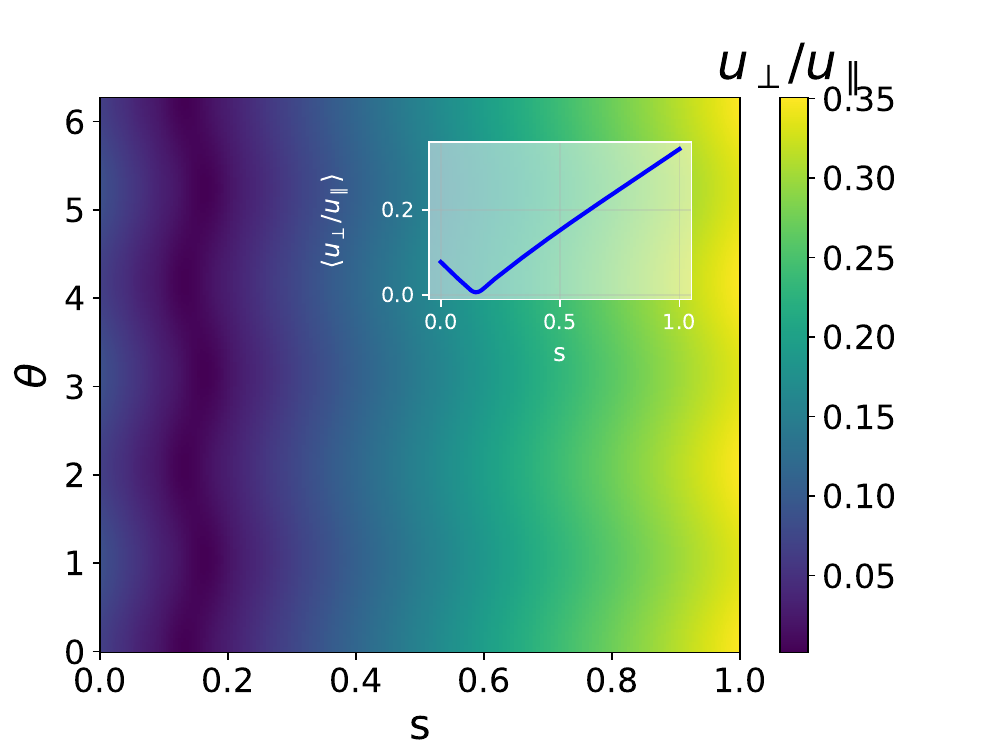}
    \put(10,70){\textbf{(d)}}
    \phantomsubcaption\label{fig:cyl_u_perp_parallel}
\end{overpic}

\caption{Example solutions in cylindrical geometry. (a) Poincaré plot of the magnetic field, (b) magnetic field norm, (c) flow norm and (d) cross-field flow ratio. Here $\delta=0.02$, $\alpha=10$, $r_-=0.5$ and $r_+=1$ used along with the parameters of \cref{tab:parameters}.}
\label{fig:cyl}
\end{figure}

Figure~\ref{fig:cyl}(\textit{a--d}) shows a solution of the model in the cylindrical geometry, using the same parameters as the HKT slab. In comparison with the slab geometry, we see in \cref{fig:cyl_poinc} that the rotational transform has a convex profile, likely because of the reduction in toroidal current cross-section due to the added curvature in the poloidal direction. As a result, the island is moved towards the interior boundary to $ s \approx 0.17$. Also, because the added curvature compressed the field lines close to the interior boundary, the norm of the magnetic field is slightly elevated in that region (\cref{fig:cyl_B_norm}). This leads to an increase in the flow norm close to the $s=0$ boundary, because the relaxed flow is proportional to the magnetic field (\cref{fig:cyl_u_norm}). The increase in the relaxed flow increases the parallel flow, and as a result, the cross-field flow ratio decreases close to the $s=0$ boundary, as seen in \cref{fig:cyl_u_perp_parallel}. In the cylindrical geometry, we can again see that the cross-field flow vanishes at the O point of the island.

\subsection{Effect of the $\alpha$ on the cross-field flow ratio, equilibrium profiles and magnetic islands}
In the slab and cylindrical geometries, the system of equations does not depend on $\omega$ and therefore the value of $\omega$ does not affect the magnetic field or pressure profiles. Although $\omega$ contributes to the toroidal flow, this contribution is evident from \cref{eq:u_z_lin}. 
The effect of the particular choice of $\alpha$, however, is not immediately obvious, because it explicitly appears in the system of equations. For this reason, in this section we set $\omega = 0$ and vary $\alpha$ to investigate how the equilibrium profiles and the flow characteristics are affected by this parameter.
The choice $\omega = 0$ is also physically motivated. From \cref{eq:u_z_lin}, if $\omega$ is interpreted as the dominant contribution to the toroidal flow and $\alpha u_\zeta^{Rx}$ is treated as a small perturbation, then $\omega = 0$ is expected in slab and cylindrical geometries. This is because, in these geometries, the toroidal direction is bounded by two walls which prevent substantial ion flow.
Additionally, in this section we restrict our investigation to the cylindrical geometry. Since the system of equations is identical for the slab and cylindrical geometries, the effects observed here apply equally to both geometries.

\begin{figure}
\centering

\begin{overpic}[width=0.49\textwidth]{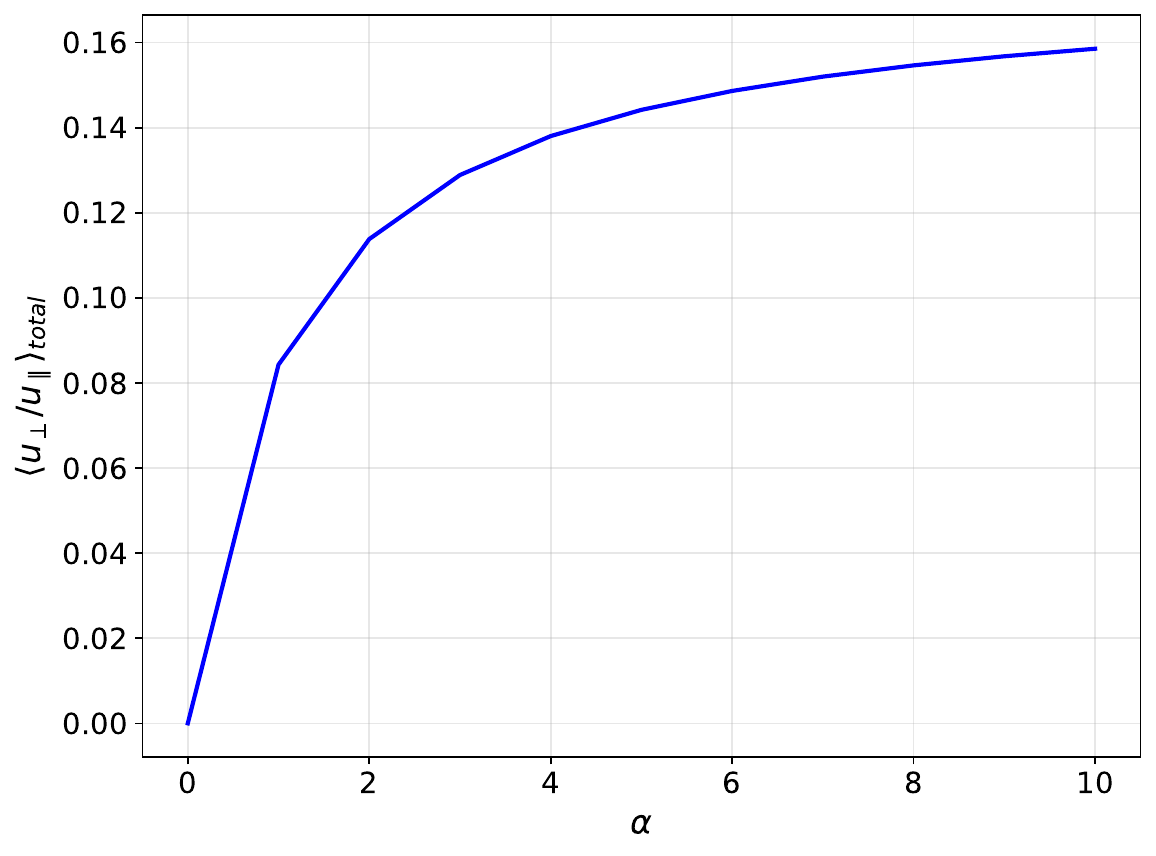}
    \put(10,70){\textbf{(a)}}
    \phantomsubcaption\label{fig:flow_anisotropy_psith_p2}
\end{overpic}
\hfill
\begin{overpic}[width=0.49\textwidth]{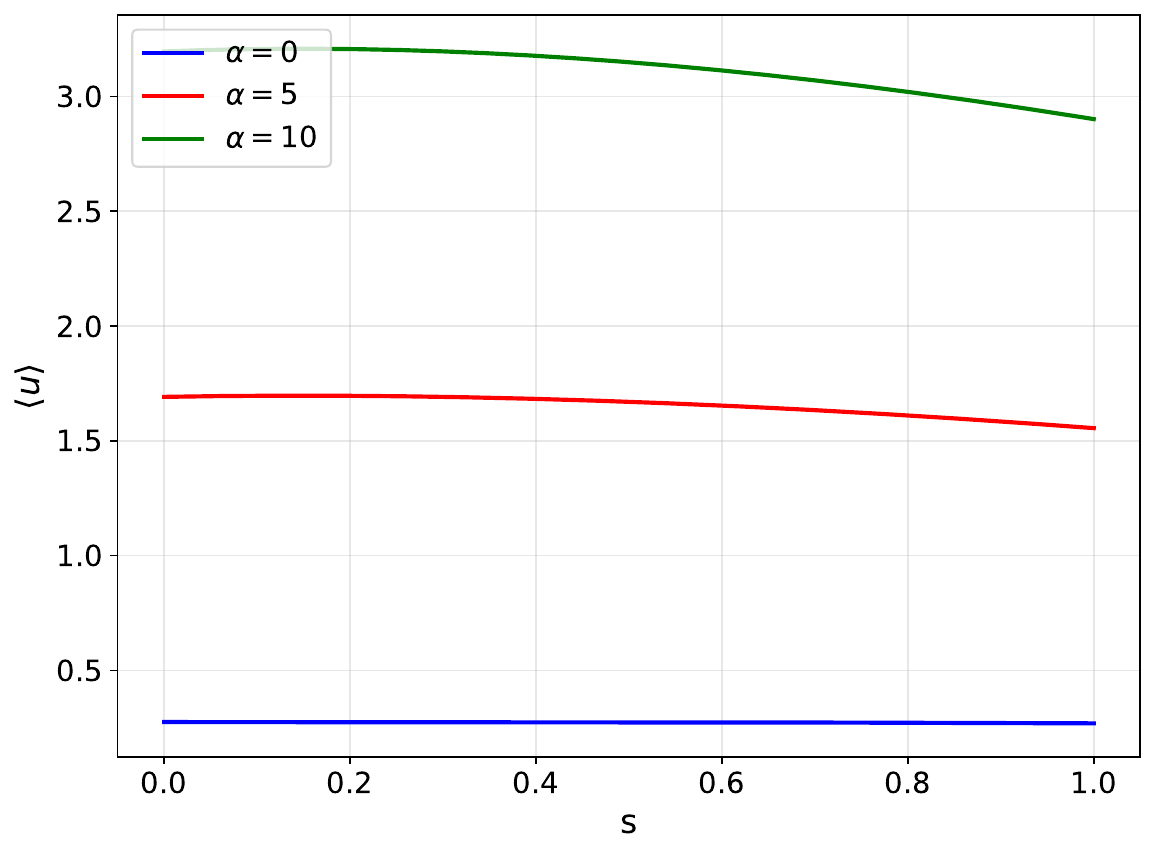}
    \put(10,70){\textbf{(b)}}
    \phantomsubcaption\label{fig:u_norm_comparison_psith_p2}
\end{overpic}

\vspace{2mm}

\begin{overpic}[width=0.49\textwidth]{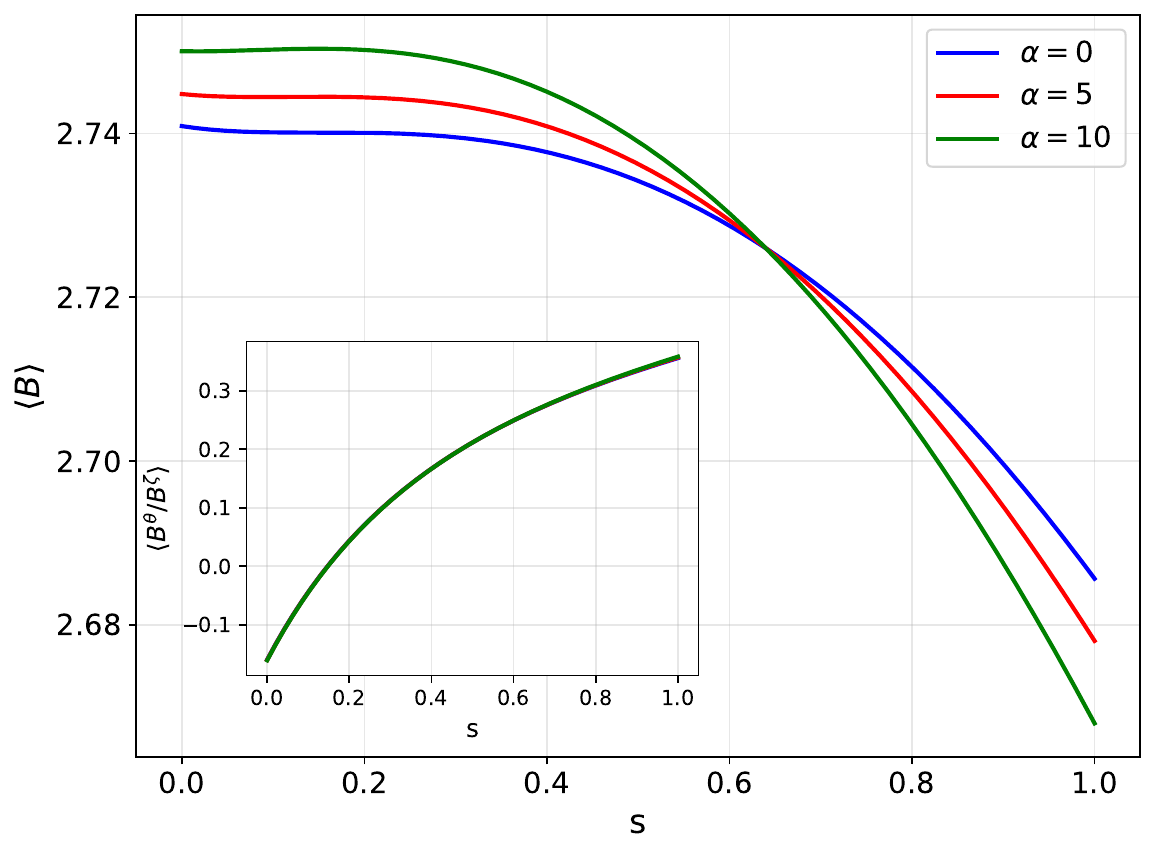}
    \put(10,70){\textbf{(c)}}
    \phantomsubcaption\label{fig:B_norm_comparison_psith_p2}
\end{overpic}
\hfill
\begin{overpic}[width=0.49\textwidth]{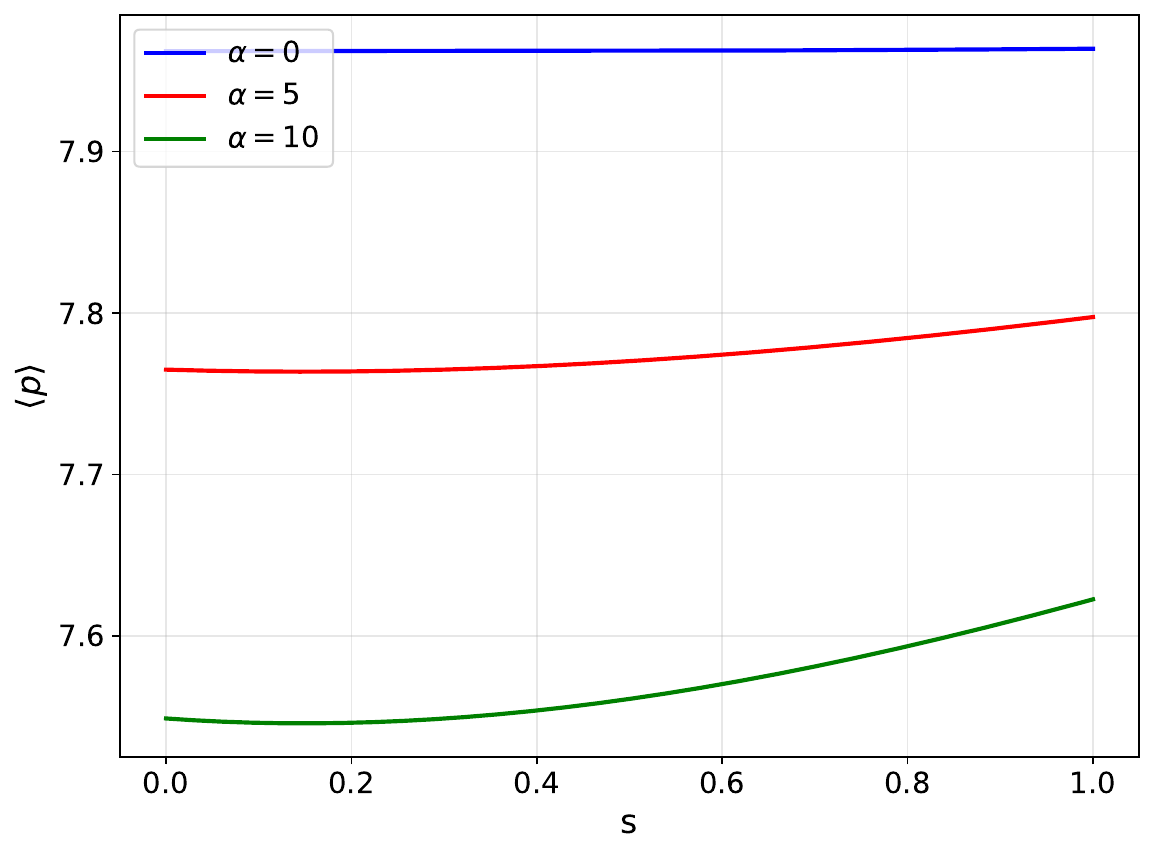}
    \put(10,70){\textbf{(d)}}
    \phantomsubcaption\label{fig:p_avg_comparison_psith_p2}
\end{overpic}

\caption{Effect of $\alpha$ on the (a) cross-field flow ratio, (b) flow norm, (c) magnetic field and (d) pressure, in cylindrical geometry.}
\label{fig:alpha}
\end{figure}

Figure~\ref{fig:flow_anisotropy_psith_p2} illustrates the cross-field flow ratio as a function of $\alpha$ in the cylindrical geometry. The cross-field flow ratio in this figure is reported as the total average of \cref{eq:flow_anis_slab} in the whole domain. As expected from \cref{eq:flow_anis_slab}, when $\alpha=0$ the cross-field flow vanishes and the cross-field flow ratio is zero, i.e.\ the flow becomes completely field-aligned. Therefore, this case is similar to the MRxMHD equilibrium with flow \citep{qu2020stepped}. In \cref{fig:u_norm_comparison_psith_p2,fig:p_avg_comparison_psith_p2} we see that in this case, the flow and pressure profiles are almost flat (though some small changes across the domain still exist). As we increase $\alpha$ from zero, the cross-field flow increases and the cross-field flow ratio grows. As this ratio increases, the change of pressure, flow and magnetic field profiles across the domain increases.  As $\alpha \rightarrow \infty$, $(u_\perp/u_\parallel)\rightarrow \abs{B_p/B_\zeta}$. The rotational transform also changes as $\alpha$ (and therefore, the cross-field flow) increases. However, from \cref{fig:B_norm_comparison_psith_p2}, one can see that this change is relatively small. Rather, the rotational transform profile is crucially affected by the poloidal and toroidal magnetic fluxes. In our numerical experiments, we also observed that the width of the islands is not significantly affected by flow parameters such as $\alpha$. It is likely because the islands are primarily determined by the domain geometry, which, because of $\vb{B}\cdot \vb{n}=0$ boundary condition, determines the geometry of magnetic field lines. As we will see, this conclusion does not hold in the geometries with a finite curvature along the $\zeta$ axis, where the flow and the geometry become strongly coupled.

It is interesting to note that, in contrast to ideal axisymmetric equilibrium codes such as FLOW \citep{guazzotto2004numerical}, the toroidal flow in our model can vanish, and flow can become exclusively poloidal. In the slab and cylindrical geometries, this can be achieved by setting $\alpha=-1$ and $\omega=0$ in \cref{eq:u_z_lin}. In such cases, \cref{eq:ber_grad_form} reduces to $\grad{h_p}=0$ or $((u^s u_s+u_\theta u^\theta)/2)+\tau\ln(\rho/\rho_\Omega)=0$, which should be solved alongside \cref{eq:belt_12,eq:belt_3}. This possibility is likely a consequence of not enforcing the ideal constraint $\curl{\qty(\vb{u}\times\vb{B})}=0$, as done in \citet{guazzotto2004numerical}, which inherently couples the poloidal direction with the toroidal direction.

\section{The toroidal geometry and Finn--Antonsen equilibrium}\label{sec:toroidal}
To construct a solution in toroidal geometry, we consider an axisymmetric system consisting of two concentric tori with a major radius $R_0$ that form the boundaries of our domain (\cref{fig:toroid_geo}). We choose the centre of the $(R,\phi,z)$ coordinate system at the centre of the torus. The inner boundary and outer boundaries,  are positioned at $\vb{\mathcal{R}}_-=(R_0+r_-\cos\theta)\hat{r}+r_-\sin \theta \hat{z}$ and $\vb{\mathcal{R}}_+=(R_0+r_+\cos\theta)\hat{r}+r_+\sin \theta \hat{z}$, respectively. The transformation equations become
\begin{subequations}
    \begin{align}
     R&=(1-s)\qty(R_0+r_-\cos \theta )+s\qty(R_0+r_+\cos \theta ),\\
     \phi&=-\zeta,\label{eq:toroid_transform_phi} \\
Z&=(1-s)r_-\sin\theta+sr_+\sin \theta,   
    \end{align}
    \label{eq:toroid_transform} 
\end{subequations}
where $r_-$ and $r_+$ are the radii of the inner and outer boundaries. In \cref{eq:toroid_transform_phi}, the negative sign ensures that the $\{s,\theta,\zeta\}$ coordinates are right-handed. One can deform the geometry by defining $r_+$ in \cref{eq:toroid_transform} as a function of $\theta$. For example, $r_+=a+\delta \cos{\theta}$, which adds a triangularity to the outer boundary of the torus. However, unlike the cylindrical geometry, due to the finite toroidal curvature of the toroidal geometry a magnetic island can exist even without such boundary perturbations. Therefore, to keep our case simple, we avoid such perturbations in this study. The metric tensor in the toroidal geometry reads
\begin{equation}
g_{ij}=\left(
\begin{array}{ccc}
\left( \frac{\partial R}{\partial s} \right)^{\!2} + \left( \frac{\partial z}{\partial s} \right)^{\!2} &
\frac{\partial R}{\partial s} \frac{\partial R}{\partial \theta} + \frac{\partial z}{\partial s} \frac{\partial z}{\partial \theta} &
0 \\[1em]
\frac{\partial R}{\partial s} \frac{\partial R}{\partial \theta} + \frac{\partial z}{\partial s} \frac{\partial z}{\partial \theta} &
\left( \frac{\partial R}{\partial \theta} \right)^{\!2} + \left( \frac{\partial z}{\partial \theta} \right)^{\!2} &
0 \\[1em]
0 & 0 & R(s,\theta)^2
\end{array}
\right)
\label{eq:toroid_metric}
\end{equation}
and therefore
\begin{equation}
    g_{33}=R^2.
    \label{eq:g_33_toroid}
\end{equation}
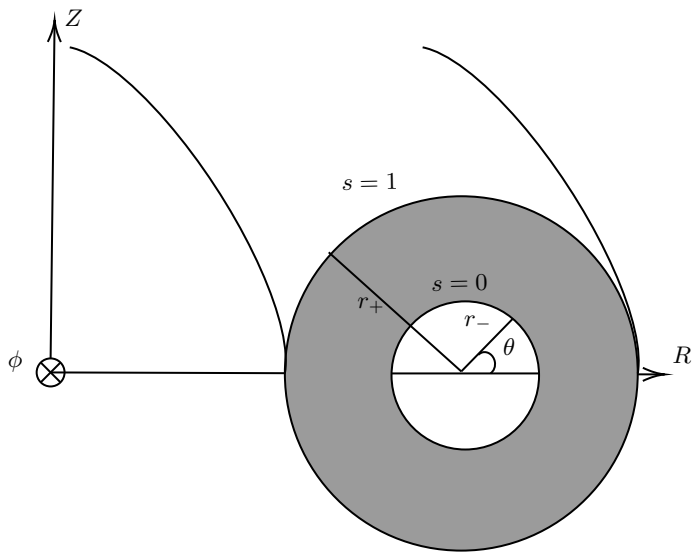
\begin{figure}
    \centering
    \input{toroid_geo.tex}
    \caption{The toroidal geometry.}
    \label{fig:toroid_geo}
\end{figure}

In toroidal geometry, the simplest way to satisfy \cref{eq:toroid_cond} is to define 
\begin{equation}
    v^\zeta=\omega,
    \label{eq:vzeta_const}
\end{equation}
where $\omega$ is a constant `angular frequency' of plasma rotation along the axis of symmetry, which here is the toroidal direction $\grad{\zeta}$. We note that \cref{eq:vzeta_const} is a special case ($\alpha=0$ case) of the \cref{eq:v_z_lin} considered in the slab and cylindrical geometries. 
Using \cref{eq:vzeta_const} in \cref{eq:ber_grad_form} one gets
\begin{equation}
    \omega u_\zeta=h_p+\frac{u^\zeta u_\zeta}{2},
    \label{eq:ber_vzeta_const_1}
\end{equation}
where an arbitrary constant is absorbed in the $\rho_\Omega$.
Using \cref{eq:vu_relation_eq}, $u_\zeta^{Rx}=g_{31}u^{s,Rx}+g_{32}u^{\theta,Rx}+g_{33}u^{\zeta,Rx}$, and after some simplifications \cref{eq:ber_vzeta_const_1} reduces to
\begin{equation}
       \frac{(u^{Rx})^2}{2}+\tau\ln\frac{\rho}{\rho_\Omega}-\frac{\omega ^2g_{33}}{2}=0.
       \label{eq:ber_vzeta_consta_2}
\end{equation}
Equation \cref{eq:ber_vzeta_consta_2} is a `generalised Bernoulli equation' that is valid as long as \cref{eq:vzeta_const} holds, in any geometry.
Interestingly, we can show that by defining the $v^\zeta$ as \cref{eq:vzeta_const}, the Finn--Antonsen equilibrium \citep{finn1983turbulent} is retrieved. In the toroidal geometry, \cref{eq:ber_vzeta_consta_2} reads
\begin{equation}
       \frac{(u^{Rx})^2}{2}+\tau\ln\frac{\rho}{\rho_\Omega}-\frac{\omega ^2R^2}{2}=0,
       \label{eq:FA_Ber}
\end{equation}
which is equivalent to (29) of \citet{finn1983turbulent}. Also, using \cref{eq:vzeta_const,eq:toroid_metric} one finds  $\vb{v}=\omega R^2 \vb{e}^\zeta=-\omega R^2\grad{\phi}$. Calculating $\curl{\vb{v}}$ and using \cref{eq:vu_relation_eq} inside \cref{eq:GenBelt_eq} we find
\begin{equation}
    \curl{(1-\frac{\nu^2}{\rho})\vb{B}}-\mu_B\vb{B}+2\nu\omega\hat{z}=0,
    \label{eq:FA_27}
\end{equation}
which is equivalent to (27) of \citet{finn1983turbulent}. Equations \cref{eq:vu_relation} and \cref{eq:isothermal_eq} are also equivalent to (25) and (26) of \citet{finn1983turbulent}. We therefore conclude that the relaxed equilibrium of \citet{finn1983turbulent} is a subclass of the new equilibria considered in this study, which is retrieved by using \cref{eq:vzeta_const} in the toroidal geometry. However, in numerical studies, we solve the reduced system of equations instead of (25)--(29) of \citet{finn1983turbulent}.

It is also instructive to make contact with the classical rotating-equilibrium theory of \citet{maschke1980exact}. Solving the generalised Bernoulli equation \cref{eq:FA_Ber} for the density gives $\rho=\rho_\Omega\exp\qty{\qty[\omega^2R^2/2-(u^{Rx})^2/2]/\tau}$, so that the density, and hence the pressure $p=\tau\rho$, acquires an explicit $\exp(\omega^2R^2/2\tau)$ dependence within each flux surface. Consequently, the pressure is not constant on a flux surface, but is centrifugally enhanced towards larger $R$. This recovers the well-known result of \citet{maschke1980exact} for axisymmetric plasmas with purely toroidal rotation; since $\tau$ is position independent in our model, the correspondence is with their case in which the temperature is a surface quantity. However, in the theory of \citet{maschke1980exact} the pressure, although not constant on a surface, is still organised by a globally defined poloidal flux function that foliates the domain into nested flux surfaces, as in ideal MHD. In the present relaxed model this foliation is not assumed: magnetic islands (and, in fully 3-D configurations, chaotic field lines) may form, and in their vicinity a global flux-surface label does not exist.

\begin{figure}
\centering

\begin{overpic}[width=0.49\textwidth]{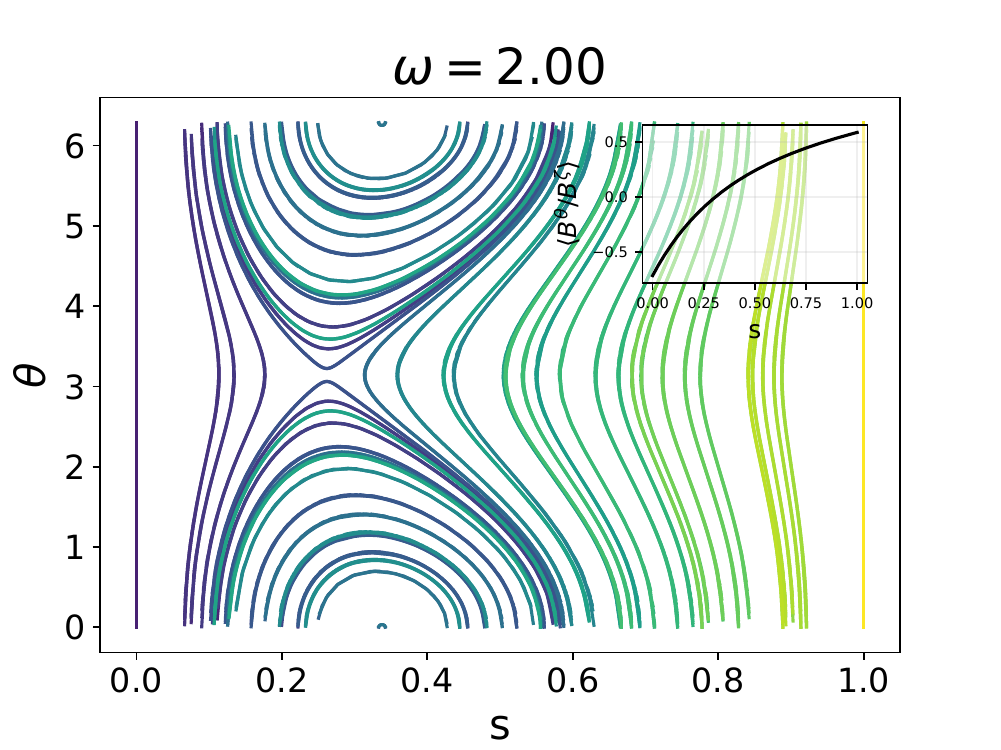}
    \put(10,70){\textbf{(a)}}
    \phantomsubcaption\label{fig:tor_poinc}
\end{overpic}
\hfill
\begin{overpic}[width=0.49\textwidth]{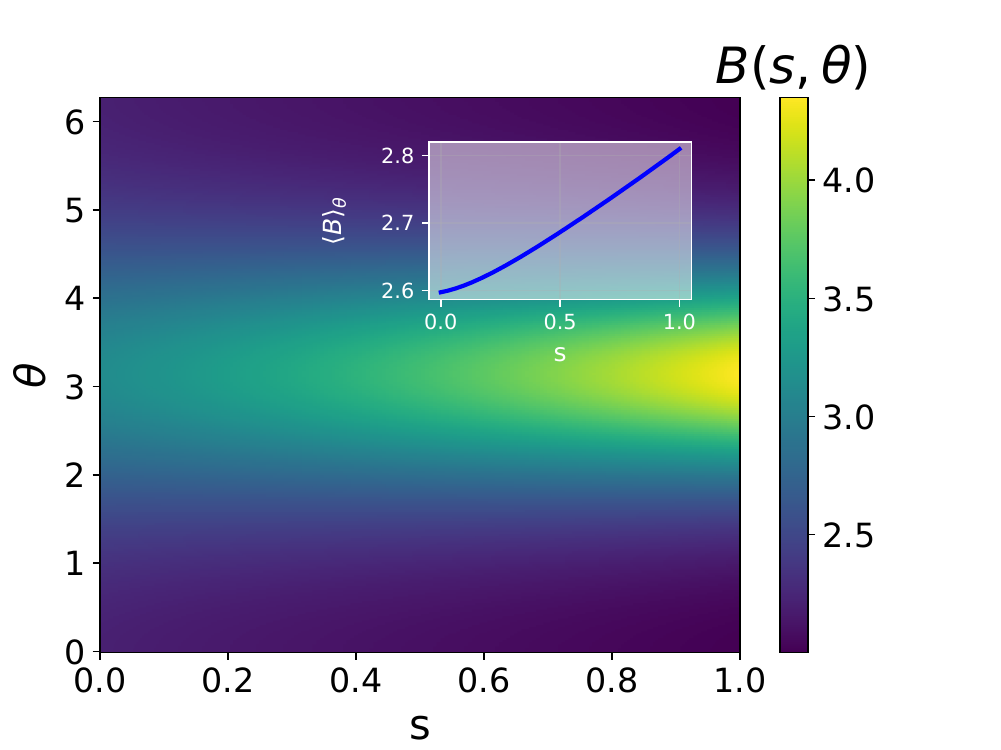}
    \put(10,70){\textbf{(b)}}
    \phantomsubcaption\label{fig:tor_B_norm}
\end{overpic}

\vspace{2mm}

\begin{overpic}[width=0.49\textwidth]{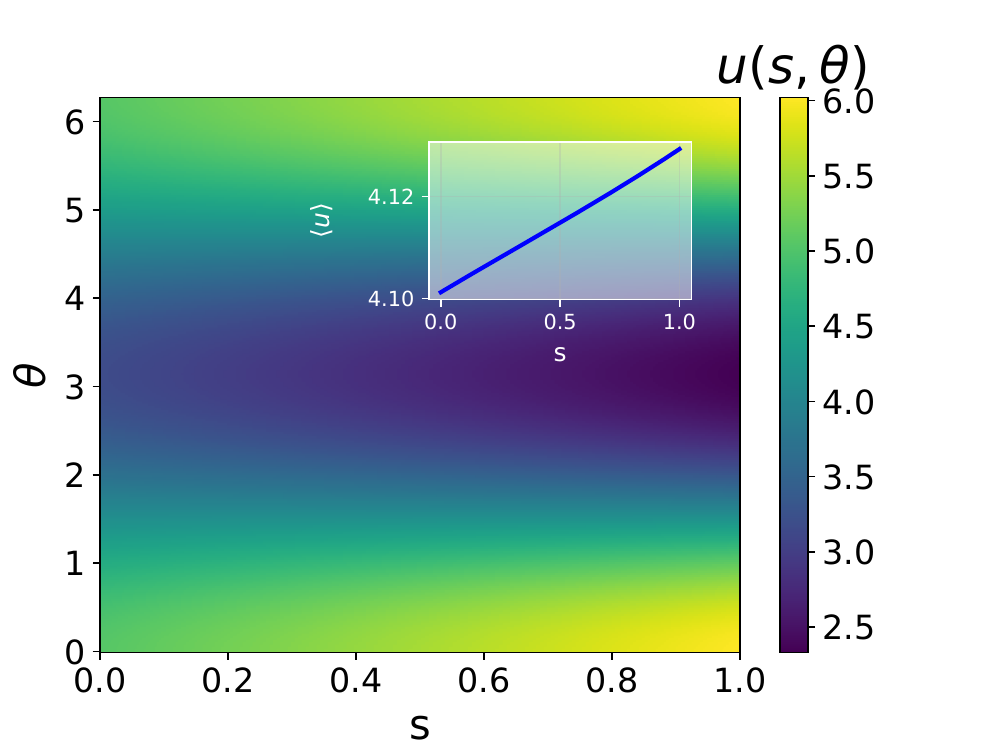}
    \put(10,70){\textbf{(c)}}
    \phantomsubcaption\label{fig:tor_u_norm}
\end{overpic}
\hfill
\begin{overpic}[width=0.49\textwidth]{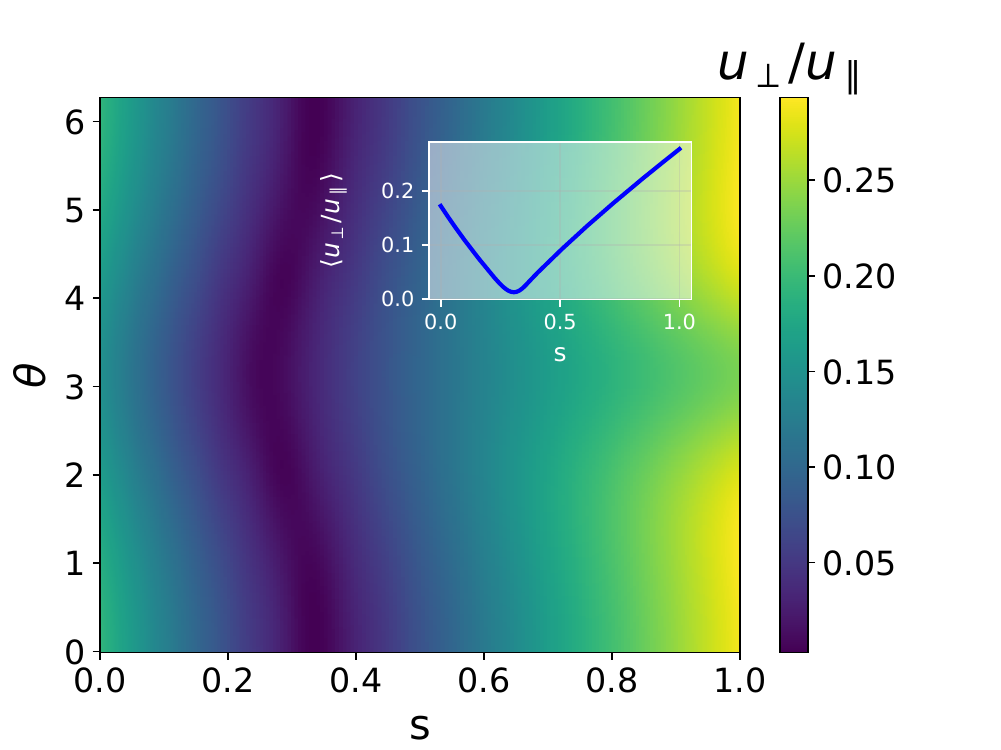}
    \put(10,70){\textbf{(d)}}
    \phantomsubcaption\label{fig:tor_u_perp_parallel}
\end{overpic}

\caption{Example solutions in toroidal geometry. (a) Poincaré plot of the magnetic field, (b) magnetic field norm, (c) flow norm and (d) cross-field flow ratio. $\delta=0$, $\omega=2$, $R_0=2$, $r_-=0.5$ and $r_+=1$ are used along with the parameters in \cref{tab:parameters}.}
\label{fig:tor}
\end{figure}

In the toroidal geometry, \cref{eq:belt_12} and \cref{eq:FA_Ber} and \cref{eq:belt_3} form the reduced system of equations that are solved for $\rho$, $\vb{A_\theta, A_\zeta}$. In addition, using \cref{eq:vzeta_const}, \cref{eq:belt_12} is reduced to
\begin{equation}
    \mu_BA_\zeta+\nu g_{33}\omega+\qty(\qty(M^{Rx}_\parallel)^2-1)B_\zeta=\mathcal{C},
    \label{eq:hel_belt_12}
\end{equation}
which has a singularity at $M^{Rx}_{\parallel}=1$, or $\nu^2=\rho$. Once again, we select the numerical values of the parameters so that this singularity is avoided.  Figure~\ref{fig:tor}(\textit{a--d}) shows the numerical solutions, in toroidal geometry, using the parameters in \cref{tab:parameters}. Figure~\ref{fig:tor_poinc} shows the Poincaré map of the magnetic field lines. One can see an island around $s\approx 0.35$, where the rotational transform vanishes. Because we are interested in the effect of the flow on the shape and width of this island, we intentionally made it large by adjusting the radii of our hollow torus. As expected, the cross-field flow, and therefore the cross-field flow ratio, vanishes in the vicinity of the island, where the flow becomes field-aligned.

\subsection{Effect of the angular frequency $\omega$ on the cross-field flow ratio and the geometry of magnetic island}

In \cref{eq:hel_belt_12,eq:ber_vzeta_consta_2}, one can see that, unlike the slab and cylindrical geometry (where $g_{33}=1$), in geometries where $g_{33}$ is not constant -- such as in helical and toroidal geometries -- the $\omega$ is not absorbed in the arbitrary constants and therefore appears in the equations. More importantly, in \cref{eq:hel_belt_12,eq:ber_vzeta_consta_2,eq:belt_Azeta}, the $\omega$ is coupled with the geometry; i.e.~it is coupled with the $g_{33}$ and $g_{32}$ components of the metric. Because of this, one expects that adjusting the cross-field flow through $\omega$ changes geometrical effects such as islands. This point will be clarified in this section.
Using \cref{eq:parallel_flow,eq:perp_flow} one can find the norm of the parallel and cross-field flows as 
\begin{subequations}
    \begin{align}
    u_\parallel&=\frac{\nu B^2+\rho\omega B_\zeta}{\rho B},\\
    u_\perp&=\frac{\abs{\omega}\sqrt{R^2B^2-B_\zeta^2}}{B},
\end{align}
\end{subequations}
and the cross-field flow ratio as
\begin{equation}
    \frac{u_\perp}{u_\parallel}=\frac{\abs{\omega} \rho\sqrt{R^2B^2-B_\zeta^2}}{\nu B^2+\rho\omega B_\zeta}.
\end{equation}
 
To demonstrate the interaction of the flow and the magnetic island, we change the angular frequency $\omega$ and solve the model in the toroidal geometry for the parameters also used in \cref{fig:tor}(\textit{a--d}). Figure~\ref{fig:tor_group}(\textit{a--g}) shows the evolution of the Poincaré plots of the magnetic field lines as $\omega$ changes from -3 to 3. Corresponding to each value of $\omega$, in \cref{fig:val_vs_omg} we have shown the width of the magnetic island, the cross-field flow ratio and the amplitude of the first and second Fourier harmonics of $B^s$. The width of the primary island is measured across the O point from separatrix to separatrix in the $s$ direction, using the Poincaré section. We note that for all values of $\omega$ in these plots, the toroidal magnetic field $B^\zeta$ remains positive, and therefore, the negative sign of $\omega$ points to the counter-toroidal-field direction. In \cref{fig:0}, we see a large island around $s\approx 0.35$ for $\omega=-3$. This island is initially located at $\theta\approx \pi$, located at the inboard of the torus in \cref{fig:toroid_geo}. In \cref{fig:4}, we see that as $\omega$ increases to $-1.25$, the width of this island shrinks. In \cref{fig:5}, we see that as we further increase $\omega$, the primary island slightly shrinks while a secondary island appears inside it. As we keep increasing $\omega$, the width of the primary island across the O point decreases, and for $\omega =-0.65$, the primary island in \cref{fig:8} disappears and only the secondary islands remain in the plot. Upon increasing $\omega$ to $-0.5$, the secondary islands move in opposite directions towards each other, and they merge at around $\theta=0$ (equivalent to $\theta=2\pi$) and recreate a primary island around that point (\cref{fig:9}). From there, by increasing the $\omega$, the secondary island disappears, and the width of the primary island grows. This behaviour of the primary and secondary islands can be explained based on the first and second harmonics of $B^s$ in \cref{fig:val_vs_omg}. In this figure, we see that as $\omega$ increases from $-3$, the width of the primary island and the value of $b^s_{m=1}$ decrease. The value of the second Fourier harmonics $b^s_{m=2}$, however, remains almost constant. Between $\omega\approx -1$ and $\omega\approx-0.4$, the second harmonic is dominant. The Poincaré plots of \cref{fig:8,fig:9} belong to this range of $\omega$, in which the primary island either disappears or is small. For $\omega \gtrsim -0.4$, the first Fourier harmonic and the island width grow. Figure~\ref{fig:val_vs_omg} also shows the correlation between the primary island width and the cross-field flow ratio ($u_\perp/u_\parallel$). The cross-field flow ratio decreases with a proportional rate to the primary island's width for $-0.65\geq \omega\geq -3$ and reaches its minimum at $\omega=-0.65$. For $\omega>-0.65$, the cross-field flow ratio grows quickly as the island width also grows. These results show that the flow, and especially the cross-field flow ratio, can significantly affect the geometry of magnetic islands and even break a primary island into a secondary island.

\begin{figure}
\centering

\begin{overpic}[width=0.49\textwidth]{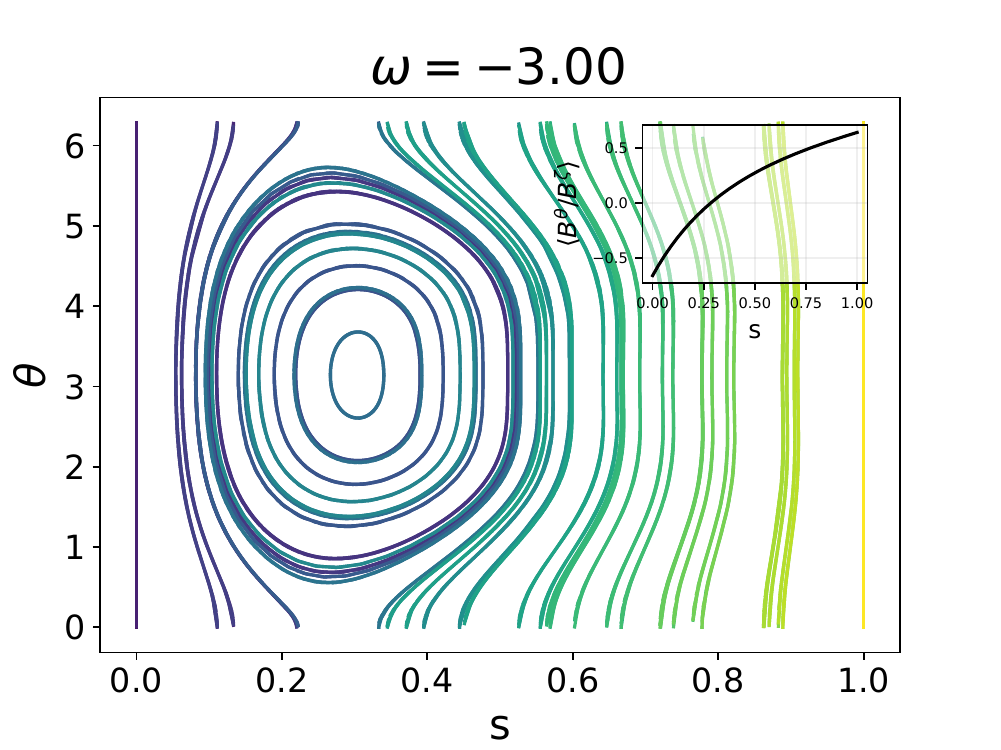}
    \put(10,70){\textbf{(a)}}
    \phantomsubcaption\label{fig:0}
\end{overpic}
\hfill
\begin{overpic}[width=0.49\textwidth]{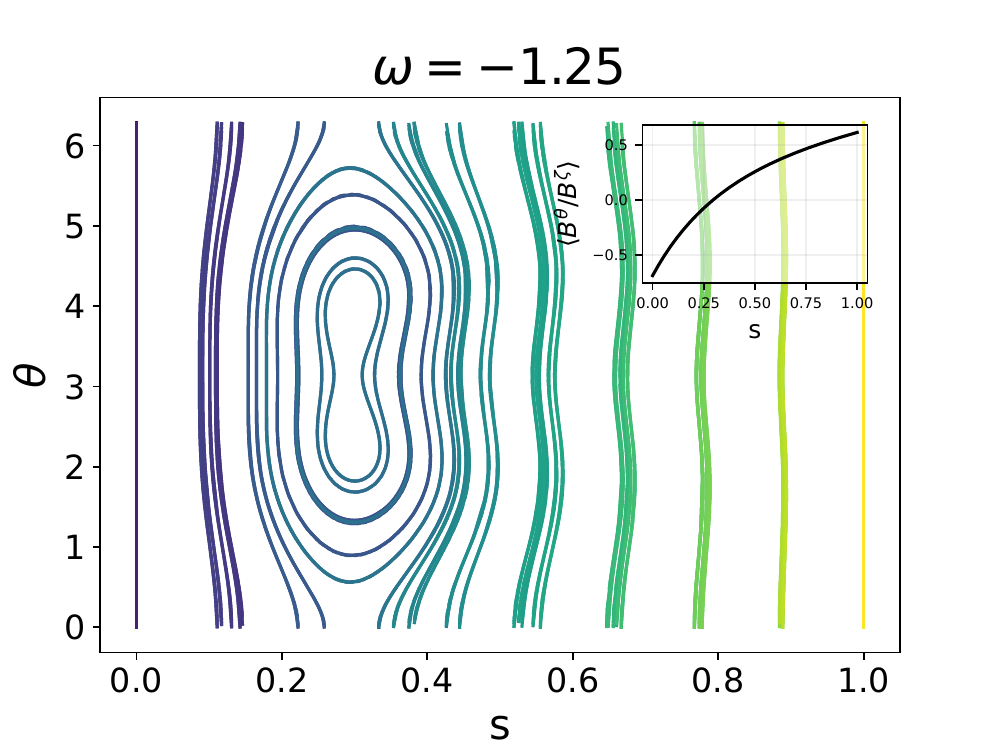}
    \put(10,70){\textbf{(b)}}
    \phantomsubcaption\label{fig:4}
\end{overpic}

\vspace{2mm}

\begin{overpic}[width=0.49\textwidth]{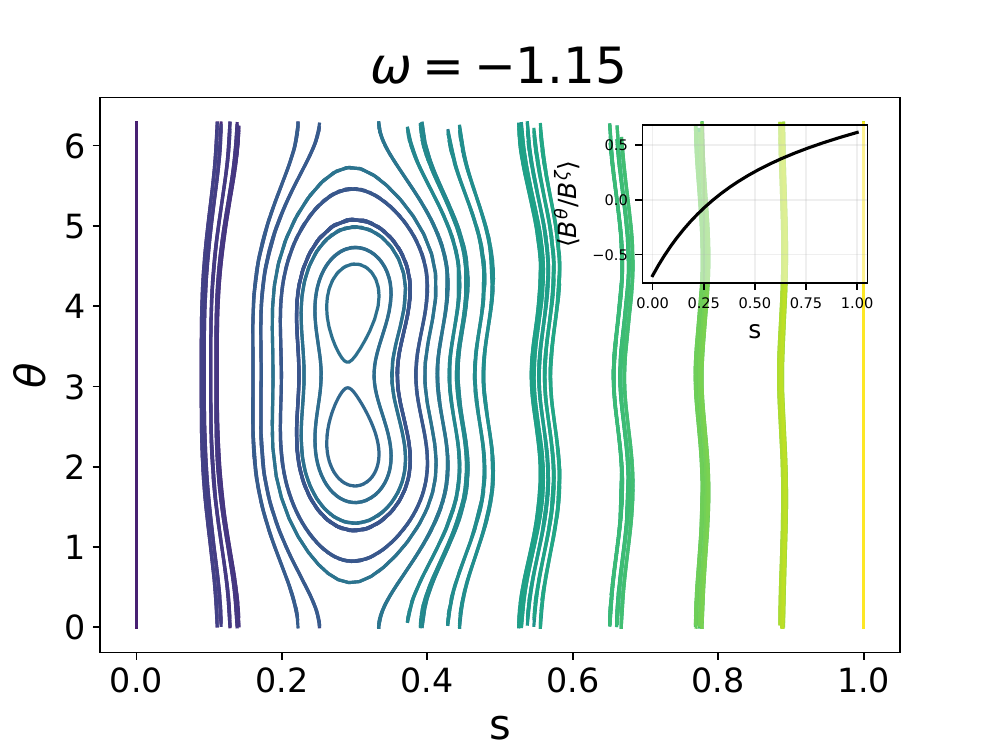}
    \put(10,70){\textbf{(c)}}
    \phantomsubcaption\label{fig:5}
\end{overpic}
\hfill
\begin{overpic}[width=0.49\textwidth]{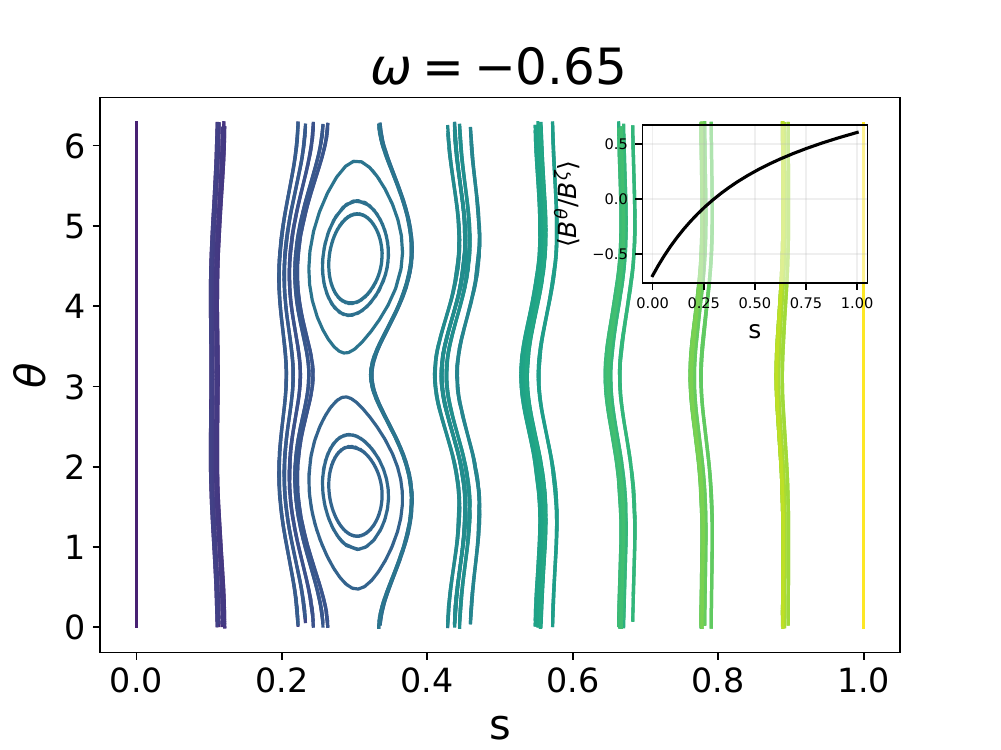}
    \put(10,70){\textbf{(d)}}
    \phantomsubcaption\label{fig:8}
\end{overpic}

\vspace{2mm}

\begin{overpic}[width=0.49\textwidth]{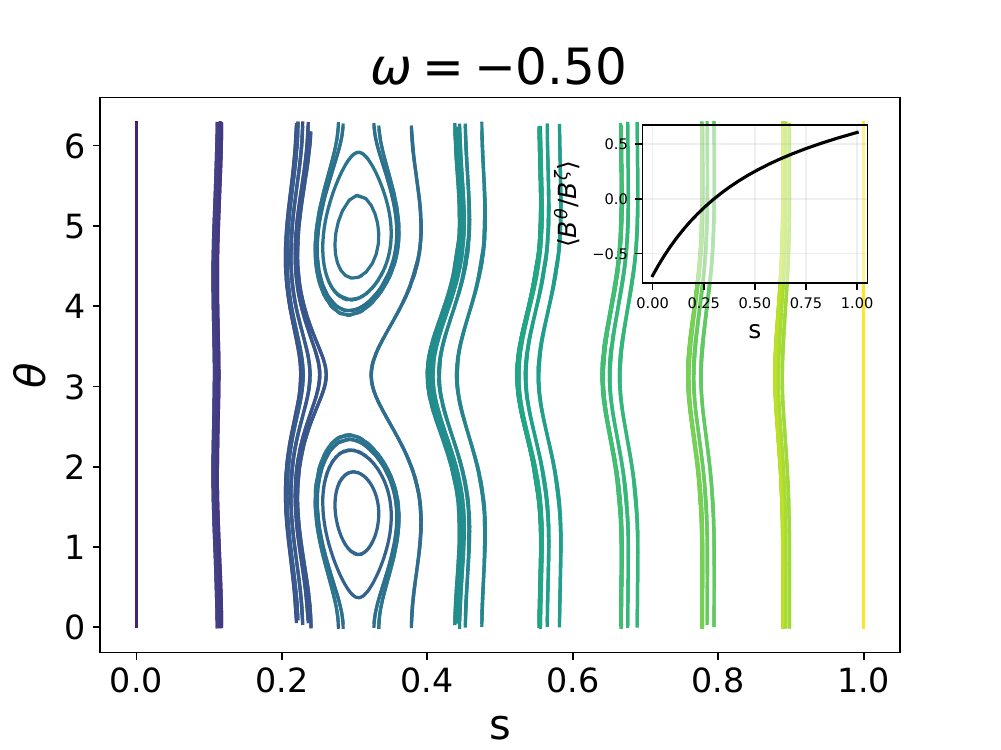}
    \put(10,70){\textbf{(e)}}
    \phantomsubcaption\label{fig:9}
\end{overpic}
\hfill
\begin{overpic}[width=0.49\textwidth]{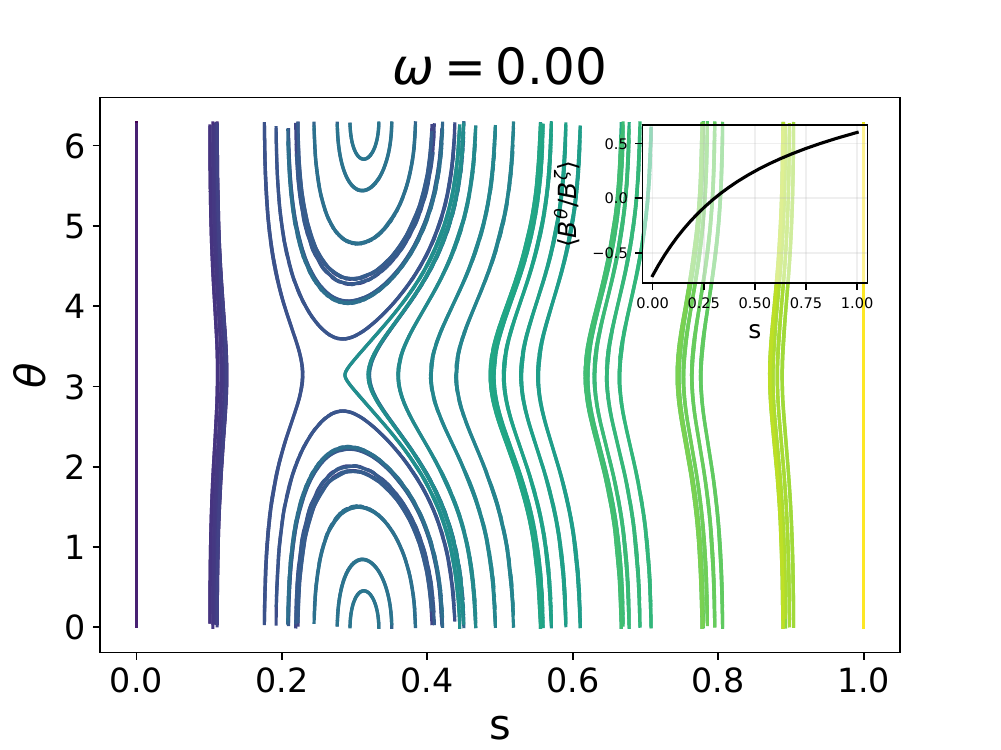}
    \put(10,70){\textbf{(f)}}
    \phantomsubcaption\label{fig:10}
\end{overpic}

\vspace{2mm}

\begin{overpic}[width=0.49\textwidth]{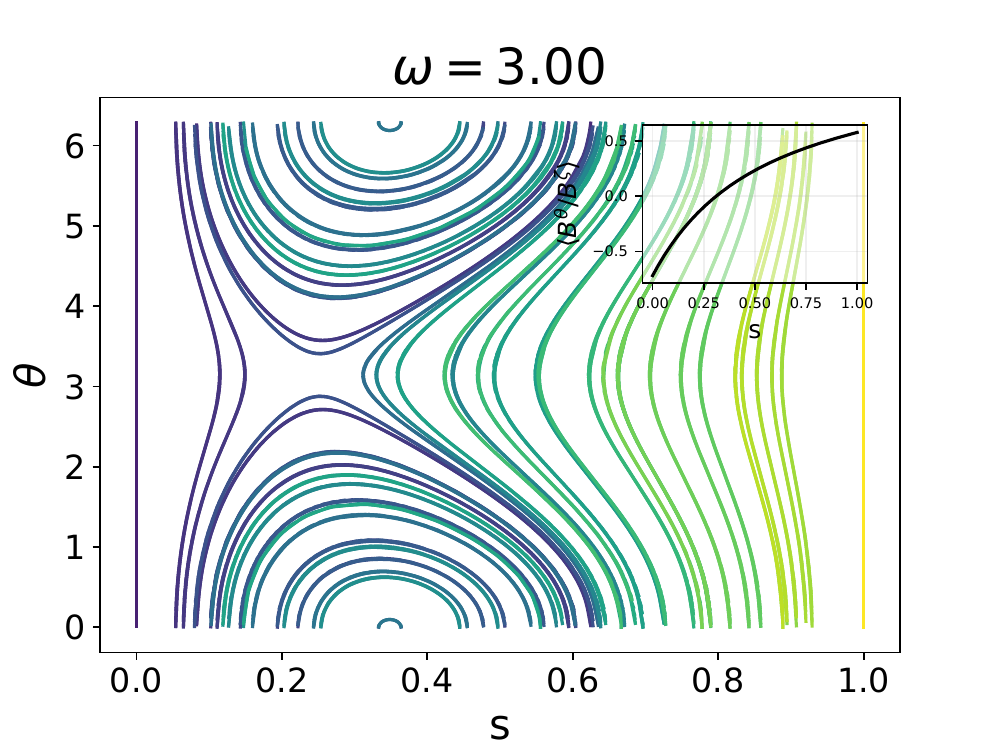}
    \put(10,70){\textbf{(g)}}
    \phantomsubcaption\label{fig:13}
\end{overpic}
\hfill
\begin{overpic}[width=0.49\textwidth]{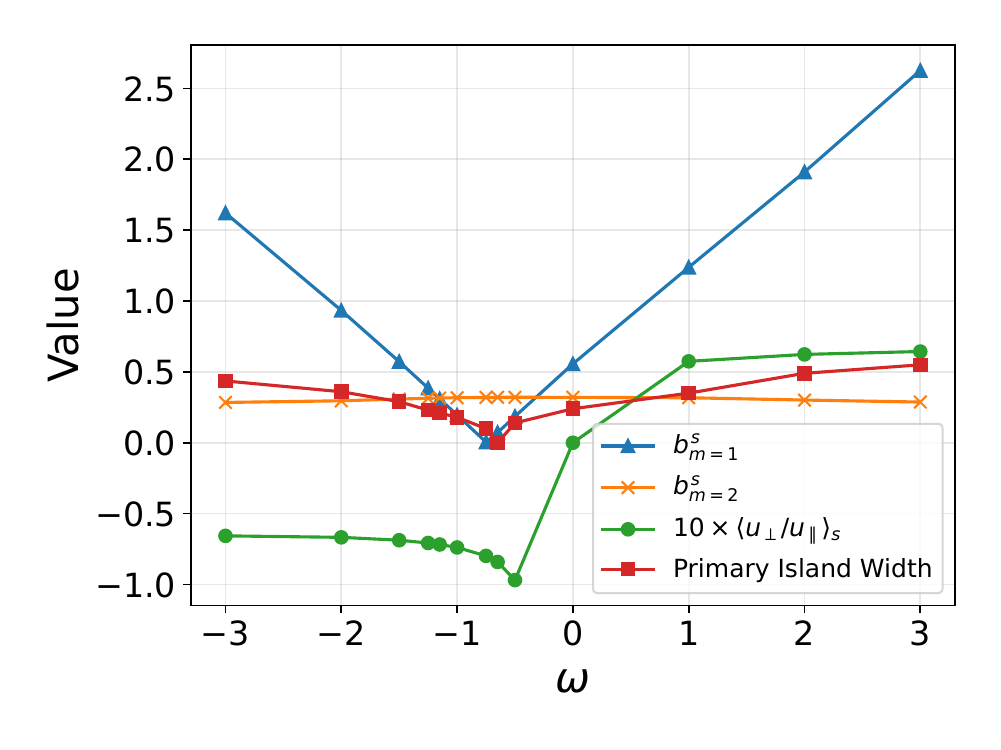}
    \put(10,70){\textbf{(h)}}
    \phantomsubcaption\label{fig:val_vs_omg}
\end{overpic}
\end{figure}

\begin{figure}\ContinuedFloat
\caption{(a--g) The evolution of primary and secondary magnetic islands versus $\omega$. (h) The primary island's width, measured across the O point in the $s$ direction, cross-field flow ratio averaged in the island region from $s=0.1$ to $s=0.6$, and the absolute value of the first and second Fourier harmonics of $B^s$.}
\label{fig:tor_group}
\end{figure}

The qualitative dependence of the island size can be captured from \cref{eq:belt_Azeta}, after some approximations. In the toroidal geometry \cref{eq:belt_Azeta} reads
\begin{gather}
  -\frac{\mu_B}{R^2}\qty(\frac{\mathcal{C}-\mu_BA_\zeta-\nu \omega R^2}{\nu^2/\rho-1})+\frac{1}{\sqrt{g}}\pdv{s}\qty[\frac{\qty(\nu^2/\rho-1)}{\sqrt{g}}\pdv{A_\zeta}{s}]\nonumber\\
  +\frac{1}{\sqrt{g}}\pdv{\theta}\qty[\frac{\qty(\nu^2/\rho-1)}{\sqrt{g}}\pdv{A_\zeta}{\theta}]=0,
    \label{eq:toroid_belt_Azeta}
\end{gather}
with $\sqrt{g}=w(sw+r_-)(R_0+(sw+r_-)\cos \theta)$ and $w\equiv r_+-r_-$. We note that the poloidal flux function, $\Psi(s,\theta)\equiv  \int_{0}^{s}\int_0^{2\pi}B^\theta(s,\theta,\zeta)\sqrt{g}\dd s\dd \zeta=-2\pi A_\zeta$, and therefore, \cref{eq:toroid_belt_Azeta} can be seen as a modified Grad--Shafranov equation written in general curvilinear coordinates. As we mentioned, we are interested in the $\qty(M^{Rx}_\parallel)^2=(\nu^2/\rho)\ll 1$ regime, and therefore \cref{eq:toroid_belt_Azeta} reads
\begin{gather}
  \frac{\mu_B}{R^2}\qty(\mathcal{C}-\mu_BA_\zeta-\nu \omega R^2)-\frac{1}{\sqrt{g}}\pdv{s}\qty(\frac{1}{\sqrt{g}}\pdv{A_\zeta}{s})
  -\frac{1}{\sqrt{g}}\pdv{\theta}\qty(\frac{1}{\sqrt{g}}\pdv{A_\zeta}{\theta})=0.
    \label{eq:toroid_belt_Azeta_lin}
\end{gather}
This equation is now linearised and decoupled from \cref{eq:FA_Ber}. To further approximate this equation, we use our numerical observation that close to the island $A_\zeta$ will always show a parabolic shape with an extremum at the O point. The term $\pdv{s}\qty(\frac{1}{\sqrt{g}}\pdv{A_\zeta}{s})$ is therefore approximated as $\mathcal{C}'/\sqrt{g}$, with $\mathcal{C}'$ being a constant found from the boundary condition \cref{eq:gauge3} ($\mathcal{C}'= -(10\psi_\theta/3\pi)$). Equation \cref{eq:toroid_belt_Azeta_lin} therefore reads
\begin{gather}
  \frac{\mu_B}{R^2}\qty(\mathcal{C}-\mu_BA_\zeta-\nu \omega R^2)-\frac{\mathcal{C}'}{(\sqrt{g})^2}
  -\frac{1}{\sqrt{g}}\pdv{\theta}\qty(\frac{1}{\sqrt{g}}\pdv{A_\zeta}{\theta})=0.
    \label{eq:toroid_belt_Azeta_island}
\end{gather}
Furthermore, we assume the aspect ratio is large ($((r_-+sw)/R_0)\ll 1$) and use expansions $1/R\approx (1/R_0)(1-\cos \theta)(1-(r_-+sw)\cos \theta/R_0)$ and $1/R^2\approx (1/R_0^2)(1-2(r_-+sw)\cos \theta/R_0)$. Around the primary island, we expect the first Fourier harmonics to be dominant and therefore, we truncate the Fourier series at the first harmonics and write $A_\zeta\approx a_{\zeta 0}(s)+a_{\zeta 1}(s)e^{i\theta}$. Substituting this relation in \cref{eq:toroid_belt_Azeta_island} and demanding that the zeroth and first Fourier harmonics of the left-hand side of this equation vanish, we get 
\begin{subequations}
\begin{align}
\frac{\mathcal{C}\mu_B}{R_0^2}
-\frac{\mathcal{C}'\left(2 R_0^2+\varrho^2\right)}{2 R_0^4 w^2 \varrho^2}
-\frac{\mu_B^2}{R_0^2}\,a_{\zeta 0}
+\left(\frac{3 r_--s w}{2 R_0^3 w^2 \varrho^2}
+\frac{\mu_B^2 \varrho}{R_0^3}\right)a_{\zeta 1}
&=\mu_B\nu\omega,\label{eq:0th_component}\\[4pt]
\frac{\dfrac{\mathcal{C}'}{w^2}-\mathcal{C}\mu_B\varrho^2}{R_0^3\varrho}
+\frac{\mu_B^2 \varrho}{R_0^3}\,a_{\zeta 0}
+\left(\frac{2 R_0^2\!\left(\varrho^{-2}-\mu_B^2 w^2\right)+1}{2 R_0^4 w^2}\right)a_{\zeta 1}
&=0,\label{eq:1th_component}
\end{align}
\end{subequations}
where $\varrho \equiv r_- + s w$. In these equations, $\omega$ only appears on the right-hand side of \cref{eq:0th_component}. Because of this, the effect of $\omega$ on the first Fourier harmonics (and therefore, the primary island's width) is due to the coupling between  $a_{\zeta 0}$ and $a_{\zeta 1}$ in \cref{eq:0th_component}. This coupling vanishes in the limit $((r_-+sw)/R_0)\rightarrow 0$, i.e.~when there is no curvature in $\zeta$ direction, and the torus reduces to a cylinder. We conclude that the effect of the flow on the $a_{\zeta 1}$ and therefore the primary island's width is due to the toroidal geometry, namely due to the curvature of the torus in the $\zeta$ direction. Solving \cref{eq:0th_component,eq:1th_component}, we can find the $a_{\zeta 0}(s)$ and $a_{\zeta 1}(s)$ coefficients and therefore $A_\zeta$.  Finally, noting that $B^s=\pdv{A_\zeta}{\theta}/\sqrt{g}$, we find the amplitude of the first Fourier harmonic of $B^s$ as
\begin{equation}
    b_{m=1}^s=\left| \frac{(r_-+s w)^3 \left(\mathcal{C}'+2 \mu_B  \nu  R_0^4 w^2 \omega \right)}{4 R_0^5 w \left(\mu_B ^2 w^2 (r_-+s w)^2-1\right)-4 \mu_B ^2
   R_0^3 w^3 (r_-+s w)^4}\right|.
   \label{eq:bs1}
\end{equation}
The dependence of the $b_{m=1}^s$ on $\omega$ is in the form of a shifted absolute value function, which is qualitatively consistent with the full numerical results in \cref{fig:val_vs_omg}. Moreover, in \cref{eq:bs1} the shift of the absolute value is proportional to the $\mathcal{C}'=-(10\psi_\theta/3\pi)$. We validated this prediction with the full numerical solutions. In numerical solutions, we observed that for positive values of $\psi_\theta$ the absolute value form of $b^s_{m=1}$ is shifted towards negative $\omega$, while for negative $\psi_\theta$ it is shifted towards the positive $\omega$. For $\psi_\theta=0$, however, $b^s_{m=1}$ at $\omega=0$ does not vanish, and a relatively small island will still exist in contradiction with \cref{eq:bs1}. This discrepancy is likely rooted in our large-aspect-ratio assumption and the first-order expansions that we used to capture the qualitative behaviour of $b_{m=1}^s$. Another effect that needs analysis at least to the second Fourier harmonic is the secondary islands and their dominance around the $-1\lesssim\omega\lesssim -0.4$ regime.

\section{Conclusion and discussion}\label{sec:conclusion}
In this work, we demonstrated several methods for solving the `semirelaxed' equilibrium model proposed by \citet{dewar2020time}. We explored the solution space of this model in different geometries and showed how the plasma flow and geometry are intrinsically coupled. In particular, we demonstrated that plasma flow, especially cross-field flow enabled by the current relaxed MHD model, can have a determining effect on equilibrium profiles and on the geometry of the $\iota = 0$ magnetic island.

In general, MHD equilibrium problems -- especially those that include plasma flow -- are underdetermined \citep{spada1992existence}, and additional structure is required to obtain physically meaningful solutions \citep{guazzotto2004numerical}. In the present work, we introduce this structure by prescribing a constrained flow of the form \cref{eq:v_troidal}. This choice is motivated by neoclassical expectations that plasma flow remains predominantly aligned with the symmetry direction. Although poloidal flow observed in some tokamak experiments can exceed predictions from neoclassical theory \citep{crombe2005poloidal,solomon2006experimental}, it remains much smaller than the toroidal flow. One of the key strengths of the present relaxed MHD model is that it permits substantial toroidal flow without entirely suppressing the poloidal component. The model also reproduces the expected local structure of island-centre flow, 
predicting vanishing cross-field velocity at magnetic-island O points in agreement 
with experimental observations \citep{estrada2016plasma,jiang2018influence}.
 It is, however, important to note that this simple RxMHD model does not directly include the neoclassical or microturbulence effects, and the consistency with those theories can only be investigated a \emph{posteriori}.

In the steady-state limit of the RxMHD equations, the assumption \cref{eq:v_troidal} yields a non-trivial \emph{solvability condition}, given by \cref{eq:toroid_cond}, which expresses a necessary compatibility between the prescribed constrained flow and the system geometry. This condition, which constitutes one of the key structural results of the present work, restricts the admissible functional form of the constrained flow and provides a systematic means of closing the steady-state equations. To satisfy this solvability condition, we choose $v^\zeta$ to be a linear function of $u_\zeta^{Rx}$ in Cartesian and cylindrical geometries, and a constant in the toroidal geometry. The latter choice naturally recovers the Finn--Antonsen equilibrium \citep{finn1983turbulent} as a special case of the present model.

By varying the flow parameters, we investigated the influence of plasma flow, particularly cross-field flow, on equilibrium profiles and on the $\iota = 0$ magnetic island. In Cartesian and cylindrical geometries, increasing the ratio of $v^\zeta$ to $u_\zeta^{Rx}$ enhances the cross-field flow and steepens the equilibrium gradients. However, the angular frequency of plasma rotation, $\omega$, does not appear explicitly in the governing equations and therefore does not affect the magnetic or pressure profiles in these geometries.

In contrast, the influence of $\omega$ is significantly more pronounced in toroidal geometry, especially with respect to the island structure. Over most of the accessible range of $\omega$, increasing its absolute value leads to a widening of the primary island. However, for small negative values of $\omega$ (corresponding to rotation counter to the toroidal field), this trend is reversed: increasing $|\omega|$ initially reduces the width of the primary island and eventually causes it to split into two secondary islands. As $|\omega|$ is increased further, these secondary islands merge, reforming a dominant primary island and restoring the original behaviour. We showed that this non-monotonic island dynamics can be explained by the behaviour of the first Fourier harmonics of $B^s$, which can be qualitatively captured using a simple first-order model. 

In contrast to ideal MHD models, the current RxMHD model allows for topological changes in the magnetic field -- such as magnetic reconnection and the formation of magnetic islands -- without explicitly introducing resistivity. Another pathway for incorporating these effects without resistivity is Voigt regularisation \citep{constantin2023magnetic,huang2025computation}. \citet{constantin2023magnetic} provides a rigorous proof that adding Voigt regularisation terms to the time-dependent ideal MHD equations does not alter the steady state. Using Dedalus, \citet{huang2025computation} investigates numerical solutions of the Voigt-MHD equations in 2-D systems. It has been found that, although Voigt regularisation can be a highly effective approach for reducing the computational expense of resistive MHD simulations, it may not resolve all singularities in the ideal limit. As a result, the ideal MHD equations remain extremely expensive to solve, and their steady states remain out of reach in that study.
We suggest a future study comparing the equilibrium solutions of the current model with those obtained from Voigt-regularised MHD. Another interesting direction would be to compare RxMHD solutions with flow against the saturation of the resistive tearing mode, similar to what was done in \citet{loizu2020direct}, \citet{loizu2023nonlinear} and \citet{balkovic2024direct} using SPEC.

An important direction for future work is the extension of the present model to fully 3-D equilibria. In modern stellarators, plasma flow is expected to align preferentially with directions of quasisymmetry in order to minimise neoclassical viscous damping \citep{helander2007rapidO}. A natural starting point in three dimensions would therefore be to prescribe the constrained flow $\vb{v}$ along an existing quasisymmetric direction, such as those associated with quasihelical or quasiaxisymmetric symmetry.
However, the solvability condition identified in this work relies fundamentally on the assumption of 2-D symmetry and does not directly generalise to fully 3-D geometries. In three dimensions, additional geometric couplings are expected to arise, and the steady-state RxMHD equations may admit multiple or no solutions for a given prescribed constrained flow. Identifying an appropriate generalisation of the solvability condition -- and determining how quasisymmetry can be incorporated in a mathematically consistent and physically meaningful way -- remains an open problem. 

Another direction for future work is the extension of the present RxMHD model to a multiregion relaxed model. In this setting, the Euler--Lagrange equations \cref{eq:equilibrium} would be solved independently within each relaxed region and coupled across ideal interfaces through the force-balance jump condition
$[[p+\tfrac{1}{2}B^2]]=0$.
In contrast to single-region equilibria, the geometry of the interfaces -- and hence the metric tensor that encodes the coordinate system -- would no longer be prescribed \emph{a priori}, but would instead emerge self-consistently as part of the solution. Developing such a multiregion RxMHD model would provide a natural pathway to incorporate cross-field flow into stepped-pressure equilibrium formulations and enable direct integration with 3-D equilibrium tools such as SPEC.

\section*{Acknowledgements}
We dedicate this study to Professor R.L. Dewar, the founder of the theoretical model discussed herein, whose pioneering contributions have profoundly shaped the development of relaxed magnetohydrodynamics. This research was supported by a grant from the Simons Foundation/SFARI (560651, AB). The authors used an AI-based language model (ChatGPT, OpenAI) to assist with formatting the manuscript (e.g. conversion between journal styles) and minor language polishing. The AI tool was not used for developing scientific ideas, deriving results or interpreting findings. All scientific content, analysis and conclusions were produced and verified solely by the authors.

Editor Per Helander thanks the referees for their advice in evaluating this article.

\section*{Declaration of interests}
The authors report no conflict of interest.

\section*{Data availability statement}
 The data that support the findings of this study are available from the corresponding author
upon reasonable request.

\bibliographystyle{jpp}
\bibliography{Refs.bib}

\end{document}

%% file: toroid_geo.tex
\tikzset{every picture/.style={line width=0.75pt}} 

\begin{tikzpicture}[x=0.75pt,y=0.75pt,yscale=-1,xscale=1]

\draw    (12.5,178) -- (14.48,11) ;
\draw [shift={(14.5,9)}, rotate = 90.68] [color={rgb, 255:red, 0; green, 0; blue, 0 }  ][line width=0.75]    (10.93,-3.29) .. controls (6.95,-1.4) and (3.31,-0.3) .. (0,0) .. controls (3.31,0.3) and (6.95,1.4) .. (10.93,3.29)   ;
\draw [color={rgb, 255:red, 0; green, 0; blue, 0 }  ,draw opacity=1 ][fill={rgb, 255:red, 0; green, 0; blue, 0 }  ,fill opacity=1 ]   (12.5,185) -- (318.5,185.99) ;
\draw [shift={(320.5,186)}, rotate = 180.19] [color={rgb, 255:red, 0; green, 0; blue, 0 }  ,draw opacity=1 ][line width=0.75]    (10.93,-3.29) .. controls (6.95,-1.4) and (3.31,-0.3) .. (0,0) .. controls (3.31,0.3) and (6.95,1.4) .. (10.93,3.29)   ;
\draw  [color={rgb, 255:red, 0; green, 0; blue, 0 }  ,draw opacity=1 ][fill={rgb, 255:red, 155; green, 155; blue, 155 }  ,fill opacity=1 ] (130,185.5) .. controls (130,136.62) and (169.62,97) .. (218.5,97) .. controls (267.38,97) and (307,136.62) .. (307,185.5) .. controls (307,234.38) and (267.38,274) .. (218.5,274) .. controls (169.62,274) and (130,234.38) .. (130,185.5) -- cycle ;
\draw  [fill={rgb, 255:red, 255; green, 255; blue, 255 }  ,fill opacity=1 ] (183.5,186.5) .. controls (183.5,166.07) and (200.07,149.5) .. (220.5,149.5) .. controls (240.93,149.5) and (257.5,166.07) .. (257.5,186.5) .. controls (257.5,206.93) and (240.93,223.5) .. (220.5,223.5) .. controls (200.07,223.5) and (183.5,206.93) .. (183.5,186.5) -- cycle ;
\draw    (218.5,184.5) -- (244.5,158) ;
\draw    (152,125) -- (218.5,184.5) ;
\draw  [draw opacity=0] (227.03,176.24) .. controls (227.63,175.6) and (228.4,175.14) .. (229.28,174.95) .. controls (231.92,174.37) and (234.6,176.38) .. (235.28,179.45) .. controls (235.82,181.93) and (234.89,184.33) .. (233.12,185.47) -- (230.5,180.5) -- cycle ; \draw   (227.03,176.24) .. controls (227.63,175.6) and (228.4,175.14) .. (229.28,174.95) .. controls (231.92,174.37) and (234.6,176.38) .. (235.28,179.45) .. controls (235.82,181.93) and (234.89,184.33) .. (233.12,185.47) ;  
\draw   (5.5,185) .. controls (5.5,181.13) and (8.63,178) .. (12.5,178) .. controls (16.37,178) and (19.5,181.13) .. (19.5,185) .. controls (19.5,188.87) and (16.37,192) .. (12.5,192) .. controls (8.63,192) and (5.5,188.87) .. (5.5,185) -- cycle ; \draw   (7.55,180.05) -- (17.45,189.95) ; \draw   (17.45,180.05) -- (7.55,189.95) ;
\draw    (183.5,185.5) -- (257.5,185.5) ;
\draw  [draw opacity=0] (199.1,22.62) .. controls (211.14,24.93) and (228.68,37.8) .. (247.85,59.93) .. controls (282.29,99.66) and (308.95,154.2) .. (307.42,181.73) .. controls (307.35,183.08) and (307.2,184.33) .. (307,185.5) -- (245.08,109.78) -- cycle ; \draw   (199.1,22.62) .. controls (211.14,24.93) and (228.68,37.8) .. (247.85,59.93) .. controls (282.29,99.66) and (308.95,154.2) .. (307.42,181.73) .. controls (307.35,183.08) and (307.2,184.33) .. (307,185.5) ;  
\draw  [draw opacity=0] (22.1,22.62) .. controls (34.14,24.93) and (51.68,37.8) .. (70.85,59.93) .. controls (105.29,99.66) and (131.95,154.2) .. (130.42,181.73) .. controls (130.35,183.08) and (130.2,184.33) .. (130,185.5) -- (68.08,109.78) -- cycle ; \draw   (22.1,22.62) .. controls (34.14,24.93) and (51.68,37.8) .. (70.85,59.93) .. controls (105.29,99.66) and (131.95,154.2) .. (130.42,181.73) .. controls (130.35,183.08) and (130.2,184.33) .. (130,185.5) ;  

\draw (165,146.4) node [anchor=north west][inner sep=0.75pt]    {$r_{+}$};
\draw (218.5,154.9) node [anchor=north west][inner sep=0.75pt]    {$r_{-}$};
\draw (237.96,166.43) node [anchor=north west][inner sep=0.75pt]  [rotate=-359.73]  {$\theta $};
\draw (156.82,84.89) node [anchor=north west][inner sep=0.75pt]  [font=\small,rotate=-358.79]  {$s=1$};
\draw (201.82,134.89) node [anchor=north west][inner sep=0.75pt]  [font=\small,rotate=-358.79]  {$s=0$};
\draw (323,170.4) node [anchor=north west][inner sep=0.75pt]    {$R$};
\draw (18,2.4) node [anchor=north west][inner sep=0.75pt]    {$Z$};
\draw (-10.5,172.4) node [anchor=north west][inner sep=0.75pt]    {$\phi $};

\end{tikzpicture}